\documentclass[a4paper,10pt]{JHEP3}
\usepackage[centertags]{amsmath}
\usepackage{amssymb}
\usepackage{graphicx}
\newcommand{\prn}[1]{\left ( #1 \right )}
\newcommand{\brk}[1]{\left [ #1 \right ]}
\newcommand{\dr}{\mathrm{d}}

\title{Anomalies and the Helicity of the Thermal State}
\author{R. Loganayagam \\
Junior Fellow, Harvard Society of Fellows, Harvard University, Cambridge, MA 02138. \\
Email:  nayagam@physics.harvard.edu
}
\abstract{We study the thermal expectation value of the following observeable
at finite temperature $T$ and chemical potential $\mu$  : 
$< \mathfrak{L}_{12} \mathfrak{L}_{34} ... \mathfrak{L}_{d-3,d-2} \mathcal{P}_{d-1} >$  where
$\mathfrak{L}_{ij}$ denote the angular
momenta, and $\mathcal{P}_i$ denotes the spatial momentum in $d$ spacetime dimensions
with $d$ even. We call this observeable the thermal helicity. 
Using a variety of arguments, we motivate the surprising assertion that thermal helicity
per unit volume is a polynomial in $T$ and $\mu$. Further, in field theories without
chiral gravitino, we conjecture that this polynomial can be derived from 
the anomaly polynomial of the  theory. We show that this conjecture is related to
the recent conjecture on gravitational anomaly induced transport made in arXiv:1201.2812 . 
We support these statements by various sphere partition function
computations in free theories.} 

\preprint{}

\begin{document}

\section{Introduction}
Since their discovery, anomalies have played a crucial role in our understanding of quantum
field theories (QFTs). Their robustness beyond perturbation theory makes them among the few powerful
tools we have in understanding strongly interacting non-supersymmetric QFTs (or for that 
matter, non-supersymmetric observeables in supersymmetric QFTs).  This is especially true
if one is interested in finite temperature/density behaviour in a strongly coupled theory.

Over the last few years, aided by various thermal field theory computations
\cite{Vilenkin:1978hb,Vilenkin:1979ui,Vilenkin:1980fu,Vilenkin:1980zv,
Vilenkin:1980ft,Vilenkin:1995um,Volovik:1999wx,Kharzeev:2009p,Loganayagam:2012pz,Gao:2012ix,2012PhRvL.109r1602S,
Golkar:2012kb,Hou:2012xg,Son:2012zy,Manes:2012hf}, 
AdS/CFT\cite{Erdmenger:2008rm,Banerjee:2008th,Torabian:2009qk,Landsteiner:2011iq,Chapman:2012my}, and general hydrodynamic arguments\cite{Son:2009tf,
Neiman:2010zi,Loganayagam:2011mu,Kharzeev:2011ds,Dubovsky:2011sk,Bhattacharya:2011tra,Banerjee:2012iz,
Jensen:2012jy,Jain:2012rh,Valle:2012em,Banerjee:2012cr,Bhattacharyya:2012xi,Jensen:2012kj}, 
much progress has been made in our understanding  of anomaly induced transport processes. 
All these studies have focused on a specific set of transport coefficients which 
could be exactly computed by linking them to the underlying
anomalies of the theory. In contrast, very little progress has been made in understanding
how anomalies control the behaviour of a system far from equilibrium. One of the primary
obstacles has been that, apriori,  we have very little intuition about what constitutes a 
suitable set of non-equilibrium observeables that are linked to anomalies. To answer
this question, it is necessary to extend, and recast these results in terms of observeables 
whose definition do not depend crucially on assumptions about thermal equilibrium. 

In this paper, we will take a first step in this direction by linking anomaly induced 
transport to a convenient observeable (namely \emph{thermal helicity}) which can in principle
be extended to non-equilibrium states.  In the rest of this introduction, we will define 
this observeable, followed by a summary of various results that constitute the bulk of this paper.

We will begin with a very simple question about the thermal state in a quantum
field theory. Take a quantum field theory living in an even dimensional
flat spacetime $\mathbb{R}^{2n-1,1}$. We will assume Poincare invariance.
To be specific, let us put a cartesian coordinate system $\prn{x^0,x^1,\ldots, x^{2n-1}}$,
and take the metric signature to be $(-++\ldots +)$.

Let us now consider a thermal state in this theory defined by a temperature 
$T$ and chemical potential(s) $\mu$ for conserved charges. This thermal state,
by definition, breaks Lorentz invariance, since it involves
the choice of a rest frame ( the thermal state is thermal with
respect to the Hamiltonian in that frame ). We will assume that this 
thermal state does not break  rotational/translational invariance 
in its rest frame. This means in particular that the 
angular-momentum of the thermal state is zero.

Let us be more explicit :  let $\mathfrak{L}_{2k-1,2k}$ denote the angular momentum in
the $(x^{2k-1},x^{2k})$ plane. Since $k=1,\ldots,(n-1)$ in $\mathbb{R}^{2n-1,1}$
we have $n-1$ such mutually commuting angular momenta. So the above statement
can be written in the form
\[ \langle \mathfrak{L}_{2k-1,2k} \rangle = 0 \]
where we have introduced the notation $\langle \ldots \rangle $ to denote the expectation
value in the thermal state. Similarly, rotational/translational invariance means 
that the spatial momentum $\vec{\mathcal{P}}$  of the thermal state is zero
\[ \langle \vec{\mathcal{P}} \rangle = 0 \]
Of course none of these symmetries forbid the thermal state from having an expectation value
for say the momentum squared - we can easily have $\langle \mathcal{P}^2 \rangle \neq 0 $ 
in the thermal state. A more interesting observeable is what we will call \emph{the helicity}
\[ \left\langle \prn{\prod_{k=1}^{n-1} \mathfrak{L}_{2k-1,2k} }\  \mathcal{P}_{2n-1} \right\rangle 
=   \langle \mathfrak{L}_{12} \mathfrak{L}_{34}\ldots \mathfrak{L}_{2n-3,2n-2} \mathcal{P}_{2n-1} \rangle
\]

The helicity of a thermal state is an interesting observeable for various reasons - one obvious reason is that
if the thermal state is parity invariant then the helicity is just zero. Hence, thermal helicity is an
interesting measure of how much the thermal state breaks parity. But as we will see in the course of this
paper, this is an interesting observeable because it seems closely linked to anomalies in the underlying
quantum theory. This statement is not at all obvious apriori, but we will marshall evidences of various kind
to show that this is indeed true.

Our main aim in this paper is to answer some of the basic questions about the helicity of the thermal state -
Is it finite ? Is it non-zero ? When is it non-zero ? What are various ways of computing the thermal 
helicity ? What are the consequences of having a thermal state with non-zero helicity ? We will answer 
these questions in the following sections one by one.

In \S\ref{sec:Zsph} we begin by exploring a very useful way to compute thermal helicity
from the thermal partition function on $S^{2n-1}\times\mathbb{R}$.  The essential idea is simple 
- define a thermal partition function on the sphere with chemical potentials
turned on for the angular momenta on the sphere. While this would be an infrared divergent
thing to do in flat spacetime\footnote{A rigid rotation in flat spacetime breaks down at 
large radius away from the centre since the velocities involved eventually exceed the relativistic
limit. We will refer the reader to \cite{Vilenkin:1980zv} for a more detailed discussion of this issue. }, 
the sphere provides a natural infrared regulator. More explicitly, 
let $L_a$ with $a=1,\ldots,n$ be the mutually commuting angular momenta on $S^{2n-1}$.
We are interested in a partition function of the form 
\begin{equation}
\mathcal{Z}[\Omega] \equiv \text{Tr}_{S^{2n-1}\times\mathbb{R}}\ \text{Exp} \brk{ -\frac{\prn{H-\mu Q-\sum_{a=1}^n \Omega_a L_a}}{T}} 
\end{equation}
As we will carefully argue in \S\ref{sec:Zsph} the flat space thermal helicity per unit spatial volume can be
obtained from the above partition function by taking the following limit 
\begin{equation}
\begin{split}
\frac{1}{\text{Vol}_{2n-1}} 
&\langle \mathfrak{L}_{12} \mathfrak{L}_{34}\ldots \mathfrak{L}_{2n-3,2n-2} \mathcal{P}_{2n-1} \rangle\\
&= \lim_{R\to \infty} \frac{T^n}{R\forall_{2n-1}}\brk{\prn{\prod_{a=1}^n\frac{\partial}{\partial \Omega_a}}
\ln \mathcal{Z}[\Omega] }_{\Omega=0} 
\end{split}
\end{equation}
where 
\[ \forall_{2n-1} \equiv \frac{2\pi^n}{(n-1)!} R^{2n-1} \]
is the volume of a sphere $S^{2n-1}$ with radius $R$. We will find it convenient to write this formula in terms of a
function 
\begin{equation}
\begin{split}
\mathfrak{F}_{anom}^\omega[T,\mu] =\lim_{R\to \infty} \brk{\prn{\prod_{a=1}^n\frac{1}{2\pi R^2}\frac{\partial}{\partial \Omega_a}}
\brk{-T\ln \mathcal{Z}[\Omega]} }_{\Omega=0} 
\end{split}
\end{equation}
so that  we can write  the thermal helicity per unit volume as 
\begin{equation}\label{eq:heliF}
\begin{split}
\frac{1}{\text{Vol}_{2n-1}} \langle \mathfrak{L}_{12} \mathfrak{L}_{34}\ldots \mathfrak{L}_{2n-3,2n-2} \mathcal{P}_{2n-1} \rangle 
= -(n-1)! (2T)^{n-1} \mathfrak{F}_{anom}^\omega[T,\mu]
\end{split}
\end{equation}
In the next section\S\ref{sec:cardy}, we consider a simple example of a 2d-CFT in the Cardy regime
and compute its thermal helicity using Cardy formula. This will allow us to draw various conclusions
about a large class of 2d-systems which we later generalise to higher dimensions.

In the next section\S\ref{sec:hydroZsph}, we turn to the physical interpretation of the function
$\mathfrak{F}_{anom}^\omega[T,\mu]$. By evaluating the partition function for a generic theory at the hydrodynamic limit, we 
will conclude that $\mathfrak{F}_{anom}^\omega[T,\mu]$ is closely linked to the chiral
vortical effect and various other transport closely linked to anomalies. In particular,
it is exactly the polynomial in $(T,\mu)$ introduced in 
\cite{Loganayagam:2011mu,Loganayagam:2012pz} to describe anomaly induced transport in 
arbitrary dimensions (see also \cite{Kharzeev:2011ds} on this regard). In particular \eqref{eq:heliF}
provides a simple way of computing the anomaly induced transport coefficients. We proceed in the subsequent sections to 
use this statement to compute these tranport coefficients in a wide variety of systems.

In section \S\ref{sec:anomFermi} we begin with a theory of free Weyl fermions in arbitrary dimensions where the transport
coefficients were already computed by the authors of \cite{Loganayagam:2012pz}
by invoking adiabaticity arguments. We give a microscopic derivation of their result
which can be formulated in terms of a remarkable relation between $\mathfrak{F}_{anom}^\omega[T,\mu]$ 
and the anomaly polynomial $\mathcal{P}_{anom} \brk{ F, \mathfrak{R}}$ of these theories 
\begin{equation}\label{eq:anomFP}
\begin{split}
\mathfrak{F}_{anom}^\omega[T,\mu] = \mathcal{P}_{anom} \brk{ F \mapsto \mu, p_1(\mathfrak{R}) \mapsto - T^2 , p_{k>1}(\mathfrak{R}) \mapsto 0 }
\end{split}
\end{equation}
where the right hand side gives a replacement rule where we replace various variables appearing in $\mathcal{P}_{anom}$
by temperature/chemical potential to get $\mathfrak{F}_{anom}^\omega$. This then gives a very interesting formula for the thermal helicity 
\begin{equation}\label{eq:heliP}
\begin{split}
\frac{1}{\text{Vol}_{2n-1}} &\langle \mathfrak{L}_{12} \mathfrak{L}_{34}\ldots \mathfrak{L}_{2n-3,2n-2} \mathcal{P}_{2n-1} \rangle\\ 
&= -(n-1)! (2T)^{n-1} \mathcal{P}_{anom} \brk{ F \mapsto \mu, p_1(\mathfrak{R}) \mapsto - T^2 , p_{k>1}(\mathfrak{R}) \mapsto 0 }
\end{split}
\end{equation}
This shows that
\begin{enumerate}
 \item Thermal helicity is a polynomial in $T$ and $\mu$.
 \item The numerical coefficients appearing in this polynomial are closely linked to underlying anomalies in the theory.
\end{enumerate}
In section \S\ref{sec:anomBose} we turn to a free theory of  chiral p-form bosons in arbitrary dimensions
whose anomaly induced transport coefficients are computed for the first time. Remarkably enough,
we find that the results in equations \eqref{eq:anomFP} and \eqref{eq:heliP} continue to hold.
In particular it follows that eqn.\eqref{eq:heliP} gives the thermal helicity of an arbitrary 
free field theory without gravitini. 

In section \S\ref{sec:anomGravitino} we then turn to the case of free chiral gravitini and their thermal
helicity. We provide both a microscopic derivation and a thermodynamic adiabaticity argument to 
conclude that the addition of gravitini spoils the relation between thermal helicity/anomalies
of the form \eqref{eq:anomFP} and \eqref{eq:heliP} (though at zero temperatures, the relation
continues to hold). We do not have an intuitive explanation of this result - but it does suggest that
the relation between finite temperature corrections in \eqref{eq:anomFP} and \eqref{eq:heliP}
and the gravitational anomalies (suggested in 4d by 
\cite{Landsteiner:2011cp,Landsteiner:2011iq,Landsteiner:2011tg,Landsteiner:2011tf}  ) is at a 
qualititatively different status to the zero temperature relation between the anomaly induced
transport and $U(1)^{n+1}$ anomalies\cite{Loganayagam:2011mu,Kharzeev:2011ds}. 

We then turn to the next simple class of theories - CFTs with gravity duals. Given the relation 
between the chiral vortical coefficient and the thermal helicity, one can directly compute the 
thermal helicity of a Blackhole using the existing computations of chiral vortical coefficient 
for various blackholes. Since all known computations in AdS/CFT satisfy \eqref{eq:anomFP} and \eqref{eq:heliP},
it leads to the natural conjecture that these relations are valid in 
any field theory with  without gravity/gravitini (or other higher spin 
fields). We conclude with discussion about further directions in \S\ref{sec:conclusion}.
This is followed by various technical appendices.

\section{Sphere Partition function and Thermal Helicity}\label{sec:Zsph}
In this section we are interested in exploring the relation between the thermal helicity of a system 
and its partition function on a sphere. To this end we begin by setting up some notation.
The next two subsections are relatively straightforward and the reader is encouraged to skim
through it for the basic definitions.

\subsection{Basic notation for $S^{2n-1}$}

Consider an even-dimensional flat euclidean space $\mathbb{R}^{2n}$ with  
cartesian co-ordinates   $\{x^1,\ldots,x^{2n}\}$. Alternately, 
we can work with the cylindrical co-ordinate systems
given by (for $a=1,2,\ldots,n$)
\begin{equation}
 \begin{split}
x^{2a-1} & = r_a \cos\phi_a\ ,\qquad x^{2a}  = r_a \sin\phi_a\\
\end{split}
\end{equation}
where   
\begin{equation}\begin{split}
r_a\geq 0\ \quad \text{and}\quad \phi_a\in[0,2\pi) 
\end{split}
\end{equation}

By fixing $\sum_{a=1}^n r_a^2 = R^2 $ or equivalently fixing $r_n = \sqrt{R^2-\sum_{i=1}^{n-1} r_i^2}$ 
one obtains the odd-dimensional sphere $S^{2n-1}$ embedded within the Euclidean space. 
The metric on the resultant $S^{2n-1}$ is given by
\begin{equation}
\begin{split}
(g)_{S^{2n-1}} &= \sum_{a=1}^n
             \brk{ dr_a^2 + r_a^2 d\phi_a^2}\\
&=  \brk{ \sum_{i,j=1}^{n-1}\prn{\delta_{ij}+\frac{r_i r_j}{r_n^2}}dr_i dr_j + \sum_{a=1}^n r_a^2 d\phi_a^2}_{r_n^2=R^2-\sum_{i=1}^{n-1}r_i^2}\\            
\end{split}
\end{equation}
We can invert this metric to get
\begin{equation}
\begin{split}
g^{-1}_{S^{2n-1} }&= \sum_{i,j=1}^{n-1}\prn{\delta_{ij}-\frac{r_i r_j}{R^2}}\partial_{r_i} \otimes \partial_{r_j}
+ \sum_{a=1}^n \frac{1}{r_a^2}\partial_{\phi_a} \otimes \partial_{\phi_a} \\
\end{split}
\end{equation}
The volume form on the sphere is
\begin{equation}\begin{split} 
d\forall  &= R d\phi_n\wedge\prod_{k=1}^{n-1} r_k dr_k \wedge d\phi_k\\
\end{split}\end{equation}
so that the total volume is given by
\begin{equation}\begin{split}\forall_{2n-1}=\int_{S^{2n-1}}  d\forall =(2\pi R)\prod_{k=1}^{n-1}
\frac{2\pi R^2}{2n-2k}= \frac{2\pi^n R^{2n-1}}{(n-1)!} \end{split}\end{equation}

Consider a 1-form on $S^{2n-1}$ of the form 
\begin{equation}\begin{split} V &\equiv \sum_{i=1}^{n-1} V_{r_i} dr_i + \sum_{a=1}^{n} V_{\phi_a} d\phi_a  
\end{split}\end{equation}
then its divergence is given by 
\begin{equation}\begin{split}\label{eq:divsph} d^\dag V &= \sum_{i,j=1}^{n-1}\brk{\delta_{ij}-\frac{r_i r_j}{R^2}}
\frac{1}{r_i}\frac{\partial}{\partial r_i}\prn{r_i  V_{r_j}} -  \frac{n}{R^2} \sum_{i=1}^{n-1}r_i  V_{r_i}
+\sum_{a=1}^n \frac{1}{r_a^2}\frac{\partial V_{\phi_a}}{\partial\phi_a}  \\
\end{split}\end{equation}

We want to now study the flat limit where $S^{2n-1}$ becomes $\mathbb{R}^{2n-1}$. To do this,
we begin by defining $z_{_{2n-1}}\equiv R \phi_n$. Then for a fixed $\{ r_i,\phi_i,z_{_{2n-1}} \}$
where $i=1,\ldots,n-1$ we take the limit $R\to \infty$. In this limit we have $r_n \approx R \to \infty $ and all the
expressions above smoothly go to their flat-space counterparts. For example, if
\begin{equation}\begin{split} V &\equiv \sum_{i=1}^{n-1} \brk{V_{r_i} dr_i + V_{\phi_i} d\phi_i } + V_{2n-1}dz_{_{2n-1}}  
\end{split}\end{equation}
then its divergence is given by 
\begin{equation}\begin{split}\label{eq:divFlat} d^\dag V &= \sum_{i}^{n-1}\brk{
\frac{1}{r_i}\frac{\partial}{\partial r_i}\prn{r_i  V_{r_i}}+ \frac{1}{r_i^2}\frac{\partial V_{\phi_i}}{\partial\phi_i} } +\frac{\partial V_{2n-1}}{\partial z_{_{2n-1}}}   \\
\end{split}\end{equation}
Note the relation between the components 
\[ V_{\phi_n} = \frac{1}{R} V_{2n-1} \]
In the next subsection, we will take up the Killing symmetries of the sphere and ask how they behave in this
flat space limit.

\subsection{The Killing vectors of $S^{2n-1}$ and the flat space limit}
The Killing vectors of $S^{2n-1}$ generate a $so(2n)$ algebra. To get an explicit expression, we begin with the 
$so(2n)$ Killing vectors of the embedding space $\mathbb{R}^{2n}$ which we write as
\begin{equation}\begin{split}
\mathfrak{u}^a_b 
&\equiv e^{i\prn{\phi_a-\phi_b}}\brk{ 
\frac{1}{2}\prn{r_a\frac{\partial}{\partial r_b}-r_b\frac{\partial}{\partial r_a} } 
-\frac{i}{2}r_ar_b \prn{\frac{1}{r_a^2}\frac{\partial}{\partial\phi_a}+\frac{1}{r_b^2}\frac{\partial}{\partial\phi_b} }
}   \\
\mathfrak{v}^{ab} 
&\equiv e^{i\prn{\phi_a+\phi_b}}\brk{ 
\frac{1}{2}\prn{r_a\frac{\partial}{\partial r_b}-r_b\frac{\partial}{\partial r_a} } 
-\frac{i}{2}r_ar_b \prn{\frac{1}{r_a^2}\frac{\partial}{\partial\phi_a}-\frac{1}{r_b^2}\frac{\partial}{\partial\phi_b} }
}   \\
\end{split}\end{equation}
These obey
\begin{equation}\begin{split}
\prn{\mathfrak{u}^a_b}^* &= - \mathfrak{u}^b_a \quad\text{and}\quad \mathfrak{v}^{ab} = - \mathfrak{v}^{ba} 
\end{split}\end{equation}
The $so(2n)$ commutation relations are
\begin{equation}\begin{split}
\brk{\mathfrak{u}^a_b\ , \mathfrak{u}^c_d} 
= \delta_b^c\mathfrak{u}^a_d-\delta^a_d\mathfrak{u}^c_b &\ ,\quad
\brk{\mathfrak{v}^*_{ab}\ ,\mathfrak{v}^{cd} } 
= \delta^{[c}_{[a}\mathfrak{u}^{d]}_{b]}\\
\brk{\mathfrak{u}^a_b\ , \mathfrak{v}^{cd}} 
= \delta_b^c\mathfrak{v}^{ad}+\delta_b^d\mathfrak{v}^{ca}&\ ,\quad 
\brk{\mathfrak{u}^a_b\ , \mathfrak{v}^*_{cd}} 
= -\delta^a_c \mathfrak{v}^*_{bd}-\delta^a_d\mathfrak{v}^*_{cb}\\
\brk{\mathfrak{v}^{ab}\ , \mathfrak{v}^{cd}} = 0 &\ ,\quad
\brk{\mathfrak{v}^*_{ab} , \mathfrak{v}^*_{cd}}= 0\\
\end{split}\end{equation}
It is evident from the commutation relations above that $\mathfrak{u}^a_b$ generate a $u(n)$ sub-algebra of 
$so(2n)$. A cartan sub-algebra is generated by the mutually commuting
Killing vectors 
\[ \mathfrak{u}^a_a\equiv -i\frac{\partial}{\partial\phi_a} \]

To get the Killing vectors of $S^{2n-1}$, we need to do the replacements
\begin{equation}\begin{split}
r_n &\mapsto \prn{R^2-\sum_{i=1}^{n-1} r_i^2 }^{\frac{1}{2}}\\
\frac{\partial}{\partial r_n} &\mapsto -\sum_{i=1}^{n-1}\frac{r_i}{r_n}  \frac{\partial}{\partial r_i}
\end{split}\end{equation}
so that we get
\begin{equation}\begin{split}
\mathfrak{u}^n_i 
&= r_n e^{i\phi_n}\frac{e^{-i\phi_i}}{2}\brk{
 \sum_{j=1}^{n-1} \prn{\delta_{ij}+\frac{r_ir_j}{r_n^2}}\frac{\partial}{\partial r_j} 
-\frac{i}{r_i}\frac{\partial}{\partial\phi_i}
- i \frac{r_i}{r_n^2}\frac{\partial}{\partial\phi_n }
}   \\
\mathfrak{v}^{ni} 
&= r_n e^{i\phi_n}\frac{e^{i\phi_i}}{2}\brk{
 \sum_{j=1}^{n-1} \prn{\delta_{ij}+\frac{r_ir_j}{r_n^2}}\frac{\partial}{\partial r_j} 
+\frac{i}{r_i}\frac{\partial}{\partial\phi_i}
- i \frac{r_i}{r_n^2}\frac{\partial}{\partial\phi_n }
}   \\
\end{split}\end{equation}

In the flat limit $R\to \infty$ with a fixed $\{r_i,\phi_i,z_{_{2n-1}}\}$ , we get
\begin{equation}\begin{split}
\mathfrak{u}^n_i  &=\prn{R+i z_{_{2n-1}}}\frac{1}{2}e^{-i\phi_i} \brk{
\frac{\partial}{\partial r_i} -\frac{i}{r_i}\frac{\partial}{\partial\phi_i}-\frac{ir_i}{R}\frac{\partial}{\partial z_{_{2n-1}}}
}+\frac{1}{R}(\ldots) \\
\mathfrak{v}^{ni} &= \prn{R+i z_{_{2n-1}}}\frac{1}{2}e^{i\phi_i} \brk{
\frac{\partial}{\partial r_i} +\frac{i}{r_i}\frac{\partial}{\partial\phi_i}-\frac{ir_i}{R}\frac{\partial}{\partial z_{_{2n-1}}}
}+\frac{1}{R}(\ldots) \\
\mathfrak{u}^n_n  &= R \frac{\partial}{\partial z_{_{2n-1}}}\\
\end{split}\end{equation}
which can be rearranged into
\begin{equation}\begin{split}
\mathfrak{u}^n_i-\mathfrak{v}^*_{ni}  &=i z_{_{2n-1}}e^{-i\phi_i} \brk{
\frac{\partial}{\partial r_i} -\frac{i}{r_i}\frac{\partial}{\partial\phi_i}}-ir_i e^{-i\phi_i} \frac{\partial}{\partial z_{_{2n-1}}}
+\frac{1}{R}(\ldots) \\
\mathfrak{u}^n_i+\mathfrak{v}^*_{ni}  &=R e^{-i\phi_i} \brk{
\frac{\partial}{\partial r_i} -\frac{i}{r_i}\frac{\partial}{\partial\phi_i}}
+\frac{1}{R}(\ldots) \\
\mathfrak{u}^n_n  &= R \frac{\partial}{\partial z_{_{2n-1}}}\\
\end{split}\end{equation}
These together with $\mathfrak{u}^i_j$ and $\mathfrak{v}^{ij}$ form the generators of $iso(2n-1)$.
Hence, we conclude that in the flat space limit, the $so(2n)$ Killing algebra of $S^{2n-1}$ 
degenates to $iso(2n-1)$ Killing algebra of $\mathbb{R}^{2n-1}$.

\subsection{Thermal Helicity and Sphere Partition function}

As advertised in the introduction, we are now ready to define a thermal partition function on the sphere
with chemical potentials turned on for the angular momenta on the sphere. Consider a QFT on
$S^{2n-1}\times \mathbb{R}$ - let $L_a$ with $a=1,\ldots,n$ be the mutually commuting angular
momenta on $S^{2n-1}$. These are the Noether charges associated with the 
Cartan Killing symmetries $-i\frac{\partial}{\partial\phi_a}$.

We are interested in a partition function of the form 
\begin{equation}
\mathcal{Z}[\Omega] \equiv \text{Tr}_{S^{2n-1}\times\mathbb{R}}\ \text{Exp} \brk{ -\frac{\prn{H-\mu Q-\sum_{a=1}^n \Omega_a L_a}}{T}} 
\end{equation}
This partition function can essentially thought of as a generating function for evaluating
various angular momentum observeables in the thermal state. In particular, using standard 
methods, it follows that 
\begin{equation}
\begin{split}
\langle L_1 L_2 \ldots L_n \rangle_{\text{connected}} = \brk{\prn{\prod_{a=1}^nT\frac{\partial}{\partial \Omega_a}}
\ln \mathcal{Z}[\Omega] }_{\Omega=0} 
\end{split}
\end{equation}
where the differentiation is done at constant $T$ and $\mu$. The subscript `connected' refers to the fact that 
all the disconnected pieces have been subtracted from the expectation value.

As a next step, let us carefully take the flat space limit of the above expression via the prescriptions 
of the previous subsections. As the flat space limit is taken, we get
\begin{equation}
\begin{split}
L_k &\mapsto \mathcal{L}_{2k-1,2k} \quad \text{for}\quad k=1,\ldots,n-1.\\
L_n &\mapsto R \mathcal{P}_{2n-1} 
\end{split}
\end{equation}
This gives the first main result of this paper
\begin{equation}
\begin{split}
\frac{1}{\text{Vol}_{2n-1}} 
\langle \mathfrak{L}_{12} \mathfrak{L}_{34}&\ldots \mathfrak{L}_{2n-3,2n-2} \mathcal{P}_{2n-1} \rangle_{\text{connected}}\\
&= \lim_{R\to \infty} \frac{T^n}{R\forall_{2n-1}}\brk{\prn{\prod_{a=1}^n\frac{\partial}{\partial \Omega_a}}
\ln \mathcal{Z}[\Omega] }_{\Omega=0} 
\end{split}
\end{equation}
where 
\[ \forall_{2n-1} \equiv \frac{2\pi^n}{(n-1)!} R^{2n-1} \]
is the volume of a sphere $S^{2n-1}$ with radius $R$. As mentioned in the introduction we will find it convenient 
to write this formula in terms of a function 
\begin{equation}
\begin{split}
\mathfrak{F}_{anom}^\omega[T,\mu] =\lim_{R\to \infty} \brk{\prn{\prod_{a=1}^n\frac{1}{2\pi R^2}\frac{\partial}{\partial \Omega_a}}
\brk{-T\ln \mathcal{Z}[\Omega]} }_{\Omega=0} 
\end{split}
\end{equation}
so that  we can write  the thermal helicity per unit volume as 
\begin{equation}\label{eq:heliF2}
\begin{split}
\frac{1}{\text{Vol}_{2n-1}} \langle \mathfrak{L}_{12} \mathfrak{L}_{34}\ldots \mathfrak{L}_{2n-3,2n-2} \mathcal{P}_{2n-1} \rangle 
= -(n-1)! (2T)^{n-1} \mathfrak{F}_{anom}^\omega[T,\mu]
\end{split}
\end{equation}
where we have removed the subscript `connected' since all the disconnected pieces automatically vanish due to the
assumed rotational invariance of the flat space thermal state.

\section{A toy example : 2d CFT at Cardy Regime}\label{sec:cardy}
Before going on to more general examples, we will take a particularly simple example
of a 2d CFT heated to a finite temperature/chemical potential. In 2d, 
the thermal `helicity' is actually just the momentum of the thermal 
state $\langle\mathcal{P}\rangle$. This can be  calculated using 
various methods - but we will adopt the method of looking at 
sphere partition functions that can be easily generalised to higher dimensions.
Our discussion in this section is standard and elementary - our primary aim being 
to phrase the well-known results in a specific form so that the reader can follow the
more involved discussion in higher-dimensions.

Consider then a 2d CFT with a $U(1)_L\times U(1)_R$ global symmetry placed 
on a circle of radius $R$. We will eventually be interested in the flat space
limit where the circle radius $R$ is taken to be very large keeping $T,\mu_{L,R}$ 
fixed. Since we are in a CFT this is same as taking the non-dimensional 
quantity $RT$ to be very large keeping $\mu_{L,R}/T$ fixed.

The partition function on the cylinder is given by 
\begin{equation}
\begin{split}
\mathcal{Z}[\Omega] &\equiv \text{Tr}_{S^1}\ \text{Exp} \brk{ -\frac{\prn{H-\mu_L Q_L -\mu_R Q_R- \Omega L}}{T}} \\
&\equiv \text{Tr}\  y^{J_0}q^{L_0-\frac{c_L}{24}}\ \tilde{y}^{\tilde{J}_0}\tilde{q}^{\tilde{L}_0-\frac{c_R}{24}}
\end{split}
\end{equation}
where we use the standard 2d CFT conventions\footnote{See for example \cite{Kraus:2006wn} with following conversion
rules :
\[(q)_{\text{Kraus}} = 2Q_L \ ,\quad (k)_{\text{Kraus}} = 4k_L \ \quad\text{and}\quad (z)_{\text{Kraus}} = \frac{\nu}{2}\]
with similar rules for the right sector.} 
\begin{equation}
\begin{split}
L_0-\frac{c_L}{24} &\equiv \frac{1}{2} \prn{HR-L}\ ,\quad \tilde{L}_0-\frac{c_R}{24} \equiv \frac{1}{2} \prn{HR+L} \\
J_0 &\equiv Q_L \ ,\quad \tilde{J}_0 \equiv Q_R \\
T_L &\equiv \frac{T}{1+R\Omega}\ ,\quad T_R \equiv \frac{T}{1-R\Omega}\\
q &\equiv e^{2\pi i\tau} \equiv e^{-\frac{1+R\Omega}{TR}} \equiv e^{-\frac{1}{T_L R}} \\
\tilde{q} &\equiv e^{-2\pi i\tilde{\tau}} \equiv e^{-\frac{1-R\Omega}{TR}} \equiv  e^{-\frac{1}{T_R R}}\\
\tau &\equiv i\frac{\prn{1+R\Omega}}{2\pi R T}\equiv \frac{i}{2\pi R T_L}\\
\tilde{\tau} &\equiv -i\frac{\prn{1-R\Omega}}{2\pi R T}\equiv -\frac{i}{2\pi R T_R}\\
y &\equiv e^{2\pi i \nu} \equiv e^{\frac{\mu_L}{T}}\ ,\quad \tilde{y}\equiv e^{-2\pi i\tilde{\nu}} \equiv e^{\frac{\mu_R}{T}}\\
\end{split}
\end{equation}

Here $H,L,Q_{L,R}$ are the Hamiltonian(energy), angular momentum and left/right charge operators respectively
while  $T,\Omega,\mu_{L,R}$ are the temperature, angular velocity and left/right chemical potentials respectively.
$L_0,\tilde{L}_0$ are the zero-mode Virasoro generators while $J_0,\tilde{J}_0$ are the zero-mode Kac-Moody generators.
The rest of the definitions for the various chiral/anti-chiral fugacities are standard.

To evaluate this partition function and the thermal helicity per unit volume we can work in the Cardy limit.
Begin with the Cardy formula for the entropy
\begin{equation}
\begin{split}
S
\approx 4\pi&\left[\quad
\sqrt{\frac{c_L}{24}\prn{\langle L_0 \rangle-\frac{c_L}{24}-\frac{1}{4k_L} \langle J_0 \rangle^2}} \right.\\
&\qquad\left.+
\sqrt{\frac{c_R}{24}\prn{\langle \tilde{L}_0 \rangle-\frac{c_R}{24}-\frac{1}{4k_R} \langle \tilde{J}_0 \rangle^2}}\quad 
\right]
\end{split}
\end{equation}
where $c_{L,R}$ are the central charges and $k_{L,R}$ are the Kac-Moody levels of the CFT under question.

Using the first law
\begin{equation}
\begin{split}
dS &= \frac{1}{T} d\langle H \rangle -\frac{\Omega}{T} d\langle L \rangle-\frac{\mu_L}{T} d\langle Q_L \rangle-\frac{\mu_R}{T} d\langle Q_R \rangle
 \\
&= \frac{1}{T_L R} d\langle L_0 \rangle  + \frac{1}{T_R R} d\langle \tilde{L}_0 \rangle-\frac{\mu_L}{T} d\langle J_0 \rangle-\frac{\mu_R}{T} d\langle \tilde{J}_0 \rangle
\end{split}
\end{equation}
we can express all the thermodynamic quatities in terms of the intensive variables 
\begin{equation}\label{eq:CFT2Thermo}
\begin{split}
S & = 4\pi \brk{\frac{c_L}{24}(2\pi R T_L)+ \frac{c_R}{24}(2\pi R T_R)} \\
&= \frac{2\pi R}{1-R^2\Omega^2}\brk{\frac{c_R+c_L}{24}(4\pi T)}+\frac{2\pi R^2\Omega}{1-R^2\Omega^2}\brk{\frac{c_R-c_L}{24}(4\pi T)} \\
\langle J_0 \rangle &\equiv \langle Q_L \rangle =  2k_L\mu_L R\prn{\frac{T_L}{T}}
=   \frac{2\pi R}{1-R^2\Omega^2}\brk{ 2k_L\prn{\frac{\mu_L}{2\pi}}}-\frac{2\pi R^2\Omega}{1-R^2\Omega^2}\brk{ 2k_L\prn{\frac{\mu_L}{2\pi}}}
\\
\langle \tilde{J}_0 \rangle &\equiv \langle Q_R \rangle =  2k_R\mu_R R\prn{\frac{T_R}{T}}
=  \frac{2\pi R}{1-R^2\Omega^2}\brk{ 2k_R\prn{\frac{\mu_R}{2\pi}}}+\frac{2\pi R^2\Omega}{1-R^2\Omega^2}\brk{ 2k_R\prn{\frac{\mu_R}{2\pi}}}
\\
\langle L_0 \rangle &-\frac{c_L}{24} 
=  (2\pi R T_L)^2\brk{\frac{c_L}{24}+k_L\prn{\frac{\mu_L}{2\pi T}}^2}
=  \prn{\frac{2\pi R }{1+R\Omega}}^2 \brk{\frac{c_L}{24}T^2+k_L\prn{\frac{\mu_L}{2\pi}}^2}
\\
\langle \tilde{L}_0\rangle&-\frac{c_R}{24}
= (2\pi R T_R)^2\brk{\frac{c_R}{24}+k_R\prn{\frac{\mu_R}{2\pi T}}^2} 
=   \prn{\frac{2\pi R }{1-R\Omega}}^2\brk{\frac{c_R}{24}T^2+k_R\prn{\frac{\mu_R}{2\pi}}^2}
\\
\langle H \rangle &= R(2\pi T_L)^2\brk{\frac{c_L}{24}+k_L\prn{\frac{\mu_L}{2\pi T}}^2}
+ R(2\pi T_R)^2\brk{\frac{c_R}{24}+ k_R\prn{\frac{\mu_R}{2\pi T}}^2} \\
&= \frac{2\pi R }{1-R^2\Omega^2}\brk{\frac{2}{1-R^2\Omega^2}-1} 
2\pi\brk{\frac{c_R+c_L}{24}T^2+k_R\prn{\frac{\mu_R}{2\pi}}^2+k_L\prn{\frac{\mu_L}{2\pi}}^2} \\
&\qquad +  \frac{2\pi R^2\Omega}{1-R^2\Omega^2}\brk{\frac{2}{1-R^2\Omega^2}} 2\pi\brk{\frac{c_R-c_L}{24}T^2
+k_R\prn{\frac{\mu_R}{2\pi}}^2-k_L\prn{\frac{\mu_L}{2\pi}}^2}\\
\langle L \rangle &= -(2\pi R T_L)^2\brk{\frac{c_L}{24}+k_L\prn{\frac{\mu_L}{2\pi T}}^2}
+ (2\pi R T_R)^2\brk{\frac{c_R}{24}+k_R\prn{\frac{\mu_R}{2\pi T}}^2} \\
&=  \frac{2\pi R }{1-R^2\Omega^2}\brk{\frac{2R^2\Omega}{1-R^2\Omega^2}} 
2\pi\brk{\frac{c_R+c_L}{24}T^2+k_R\prn{\frac{\mu_R}{2\pi}}^2+k_L\prn{\frac{\mu_L}{2\pi}}^2} \\
&\qquad  + \brk{\frac{\Omega^{-1}+R^2\Omega}{1-R^2\Omega^2}}  \frac{2\pi R^2\Omega }{1-R^2\Omega^2}  2\pi\brk{\frac{c_R-c_L}{24}T^2
+k_R\prn{\frac{\mu_R}{2\pi}}^2-k_L\prn{\frac{\mu_L}{2\pi}}^2}\\
\end{split}
\end{equation}

Without much ado we can now directly compute the momentum density in the flat space thermal state. We do this
by taking the angular momentum density on the cylinder at zero angular velocity, divide by radius and then 
take radius to infinity : we get
\begin{equation}
\begin{split}
\frac{1}{\text{Vol}} \langle \mathcal{P} \rangle 
= \lim_{R\to \infty} \brk{\frac{1}{R\forall_{S^1}}\langle L \rangle  }_{\Omega=0} =  2\pi\brk{\frac{c_R-c_L}{24}T^2
+k_R\prn{\frac{\mu_R}{2\pi}}^2-k_L\prn{\frac{\mu_L}{2\pi}}^2}
\end{split}
\end{equation}
This is the expression for the thermal helicity of a 2d CFT that we were after. 

Compare this  against the  anomaly polynomial of this 2d CFT 
\begin{equation}\label{eq:2dCFTP}
\mathcal{P}_{anom}[F,\mathfrak{R}] 
= -2\pi\brk{
k_R\prn{\frac{F_R}{2\pi}}^2
-k_L\prn{\frac{F_L}{2\pi}}^2
- p_{_1}(\mathfrak{R}) \frac{c_R-c_L}{24}}
\end{equation}
where the $1^{st}$-Pontryagin class denoted by $p_{_1}(\mathfrak{R})$ which is a $4$-form 
defined via
\[ p_{_1}(\mathfrak{R}) \equiv -\frac{1}{2(2\pi)^2} \mathfrak{R}_{a}{}^{b}\wedge
\mathfrak{R}_{b}{}^{a} \]
where $\mathfrak{R}_{ab}$ represents the curvature 2-forms. Comparing these 
expressions, we get a remarkable replacement rule for the thermal helicity(momentum)
per unit volume(length) of a 2d CFT
\begin{equation}
\begin{split}
\frac{1}{\text{Vol}} \langle \mathcal{P} \rangle  &= -\mathcal{P}_{anom}[F_R\mapsto \mu_R, F_L\mapsto \mu_L,  p_{_1}(\mathfrak{R})\mapsto -T^2]
\end{split}
\end{equation}

While we are essentially done, for comparison with later sections let us
rederive this result in terms of the partition function of the cylinder.
The cylider partition function at large $R$ is dominated by the thermal 
state and hence can be approximated in terms of the thermodynamic quantities 
\begin{equation}
\begin{split}
\ln \mathcal{Z}[\Omega]
&\approx S-\frac{\langle H-\mu_L Q_L -\mu_R Q_R- \Omega L\rangle}{T} 
\end{split}
\end{equation}
After subtituting for these quatities using \eqref{eq:CFT2Thermo}, we get\footnote{
Equivalently we can directly take the following expression from \cite{Kraus:2006wn}
\begin{equation}
\begin{split}
\ln \mathcal{Z}[\Omega]
&\approx 2\pi i \brk{\frac{c_L}{24\tau}-\frac{k_L\nu^2}{\tau}}-2\pi i \brk{\frac{c_R}{24\tilde{\tau}}-\frac{k_R\tilde{\nu}^2}{\tilde{\tau}}}\\
\end{split}
\end{equation}
which is valid in the Cardy regime.
}
\begin{equation}
\begin{split}
\ln \mathcal{Z}[\Omega]
&= (2\pi R) 
2\pi\brk{\frac{c_R T_R}{24}+\frac{c_L T_L}{24}+k_R T_R\prn{\frac{\mu_R}{2\pi T}}^2+k_LT_L\prn{\frac{\mu_L}{2\pi T}}^2}\\
&= \frac{1}{T}\left[\frac{2\pi R }{1-R^2\Omega^2}
2\pi\brk{\frac{c_R+c_L}{24}T^2+k_R\prn{\frac{\mu_R}{2\pi}}^2+k_L\prn{\frac{\mu_L}{2\pi}}^2} \right. \\
&\qquad \left.+  \frac{2\pi R^2\Omega}{1-R^2\Omega^2} 2\pi\brk{\frac{c_R-c_L}{24}T^2
+k_R\prn{\frac{\mu_R}{2\pi}}^2-k_L\prn{\frac{\mu_L}{2\pi}}^2} \right]\\ 
\end{split}
\end{equation}
We can now calculate the function $\mathfrak{F}_{anom}^\omega$ and relate it to the thermal helicity
(as deduced in equation \eqref{eq:heliF} before) and the anomaly polynomial 
\begin{equation}
\begin{split}
\mathfrak{F}_{anom}^\omega &\equiv \lim_{R\to \infty} \brk{\frac{1}{2\pi R^2}\frac{\partial}{\partial \Omega}
\brk{-T\ln \mathcal{Z}[\Omega]}}_{\Omega=0}\\
&=-2\pi\brk{\frac{c_R-c_L}{24}T^2
+k_R\prn{\frac{\mu_R}{2\pi}}^2-k_L\prn{\frac{\mu_L}{2\pi}}^2}\\
&= \mathcal{P}_{anom}[F_R\mapsto \mu_R, F_L\mapsto \mu_L,  p_{_1}(\mathfrak{R})\mapsto -T^2]\\
&= - \frac{1}{\text{Vol}} \langle \mathcal{P} \rangle  \\
\end{split}
\end{equation}
Hence, $\mathfrak{F}_{anom}^\omega$ can be obtained from the anomaly polynomial $\mathcal{P}_{anom}$
via a simple substitution rule as advertised - in fact,  this justifies the subscript `anom' 
in $\mathfrak{F}_{anom}^\omega$. 

Note that since this result essentially follows from assuming Cardy-like growth, we have
calculated the thermal helicity of a very diverse variety of 2d CFTs and have linked 
it to their anomaly polynomial. This generalises to a wide class of theories 
the replacement rule found by authors of \cite{Loganayagam:2012pz}\footnote{
One of the systems studied in  \cite{Loganayagam:2012pz} was a free theory of 2d
Weyl fermions with various charges under 
$F_L= F_R = F$ which result in anomaly coefficients 
\[k_R-k_L = \frac{1}{2}\sum_{species} \chi_{_{d=2}} q^2 \qquad\text{and}\qquad 
c_R - c_L = \sum_{species} \chi_{_{d=2}} \]
where $ \chi_{_{d=2}}$ is the 2d-chirality. We will revisit this system and its 
higher dimensional counterparts in section \S\ref{sec:anomFermi}.
}

Given the various examples of AdS$_3$/CFT$_2$ along with the well-known statement that 
BTZ thermodynamics is almost always captured by Cardy formula, it follows that 
the formulae of this section are valid for all the strongly coupled theories 
dual to AdS$_3$ provided they obey Cardy formula.

Our essential aim in this article is to generalise this exercise to various
higher dimensional examples. Before we proceed to do that, let us make a useful observation - using the fact that 
the Cardy pressure (i.e., pressure of the thermal CFT in the Cardy limit) is given by
\begin{equation}\label{eq:pCardy}
\begin{split}
p &\approx 2\pi\brk{\frac{c_R+c_L}{24}T^2+k_R\prn{\frac{\mu_R}{2\pi}}^2+k_L\prn{\frac{\mu_L}{2\pi}}^2}
\end{split}
\end{equation}
the Cardy approximation can be phrased as 
\begin{equation}
\begin{split}
\ln \mathcal{Z}[\Omega] 
&\approx \frac{1}{T}\brk{p\frac{\forall_{S^1}}{1-R^2\Omega^2}- \mathfrak{F}_{anom}^\omega \frac{2\pi R^2\Omega}{1-R^2\Omega^2} }\\
S &= 
\frac{\partial p}{\partial T} \frac{\forall_{S^1}}{1-R^2\Omega^2}
-\frac{\partial \mathfrak{F}_{anom}^\omega }{\partial T}\frac{2\pi R^2\Omega}{1-R^2\Omega^2}\\
\langle Q_L \rangle &=
\frac{\partial p}{\partial \mu_L} \frac{\forall_{S^1}}{1-R^2\Omega^2}
-\frac{\partial \mathfrak{F}_{anom}^\omega }{\partial \mu_L}\frac{2\pi R^2\Omega}{1-R^2\Omega^2}
\\
\langle Q_R \rangle &=
\frac{\partial p}{\partial \mu_R} \frac{\forall_{S^1}}{1-R^2\Omega^2}
-\frac{\partial \mathfrak{F}_{anom}^\omega }{\partial \mu_R}\frac{2\pi R^2\Omega}{1-R^2\Omega^2}
\\
\langle H \rangle &=
p\frac{\forall_{S^1} }{1-R^2\Omega^2}\brk{\frac{2}{1-R^2\Omega^2}-1} 
- \mathfrak{F}_{anom}^\omega \brk{\frac{2}{1-R^2\Omega^2}}\frac{2\pi R^2\Omega}{1-R^2\Omega^2} \\
\langle L \rangle &= p \frac{\forall_{S^1} }{1-R^2\Omega^2}\brk{\frac{2R^2\Omega}{1-R^2\Omega^2}} 
- \mathfrak{F}_{anom}^\omega   \brk{\frac{\Omega^{-1}+R^2\Omega}{1-R^2\Omega^2}}\frac{2\pi R^2\Omega }{1-R^2\Omega^2} \\
\end{split}
\end{equation}
where $p$ is the Cardy-pressure given by \eqref{eq:pCardy} and 
$\mathfrak{F}_{anom}^\omega$ is related to the thermal helicity on one hand and the anomaly polynomial on the other hand via
\begin{equation}
\begin{split}
\mathfrak{F}_{anom}^\omega &= \mathcal{P}_{anom}[F_R\mapsto \mu_R, F_L\mapsto \mu_L,  p_{_1}(\mathfrak{R})\mapsto -T^2]= - \frac{1}{\text{Vol}} \langle \mathcal{P} \rangle  \\
\end{split}
\end{equation}
The significance of these forms is that these can be readily generalised to higher dimensions as we will argue in the next section.

In particular, in the subsequent sections we would like to argue that the partition function
of higher dimensional CFTs with the space time dimensions $d=2n$  takes the form
\begin{equation}\label{eq:higherdim}
\begin{split}
\ln \mathcal{Z}[\Omega] 
&\approx \frac{1}{T}\left[p\frac{\forall_{S^{2n-1}}}{\prod_{b=1}^n (1-R^2\Omega_b^2)}+\ldots\right.\\
&\qquad\left.- \mathfrak{F}_{anom}^\omega \prod_{b=1}^n\frac{2\pi R^2\Omega_b }{1-R^2\Omega_b^2}+\ldots\right]\\
\end{split}
\end{equation}
where $p$ is again the pressure  and $\mathfrak{F}_{anom}^\omega$ is a quantity related to 
the thermal helicity on one hand and the anomaly polynomial on the other hand. 

Other thermodynamic quantites follow from this partition function
\begin{equation}
\begin{split}
S &= 
\frac{\partial p}{\partial T} \frac{\forall_{S^{2n-1}} }{\prod_{b=1}^n (1-R^2\Omega_b^2)}+\ldots\\
&\qquad-\frac{\partial \mathfrak{F}_{anom}^\omega }{\partial T}\prod_{b=1}^n\frac{2\pi R^2\Omega_b }{1-R^2\Omega_b^2}+\ldots\\
\langle Q \rangle &=
\frac{\partial p}{\partial \mu} \frac{\forall_{S^{2n-1}} }{\prod_{b=1}^n (1-R^2\Omega_b^2)}+\ldots\\
&\qquad-\frac{\partial \mathfrak{F}_{anom}^\omega }{\partial \mu}\prod_{b=1}^n\frac{2\pi R^2\Omega_b }{1-R^2\Omega_b^2}
+\ldots\\
\langle H \rangle &=
p\frac{\forall_{S^{2n-1}} }{\prod_{b=1}^n (1-R^2\Omega_b^2)}\brk{\sum_{a=1}^n\frac{2}{1-R^2\Omega_a^2}-1}+\ldots\\
&\qquad- \mathfrak{F}_{anom}^\omega \brk{\sum_{a=1}^n\frac{2}{1-R^2\Omega_a^2}}\prod_{b=1}^n\frac{2\pi R^2\Omega_b }{1-R^2\Omega_b^2} +\ldots\\
\langle L_a \rangle &= p \frac{\forall_{S^{2n-1}} }{\prod_{b=1}^n (1-R^2\Omega_b^2)}\brk{\frac{2R^2\Omega_a}{1-R^2\Omega_a^2}} +\ldots\\
&\qquad- \mathfrak{F}_{anom}^\omega   \brk{\frac{\Omega_a^{-1}+R^2\Omega_a}{1-R^2\Omega_a^2}}\prod_{b=1}^n\frac{2\pi R^2\Omega_b }{1-R^2\Omega_b^2} +\ldots\\
\end{split}
\end{equation}

\section{Thermal Helicity and Hydrodynamics}\label{sec:hydroZsph}
In higher dimensions, we do not have powerful tools like modular invariance. Hence, the reader
might wonder what we could possibly be said about the sphere partition functions/ thermal helicity of
generic theories in higher dimensions. We will tackle this question in this section before 
turning to some simple examples in the rest of the paper.

Our quest is to approximate sphere partition functions in the limit where the radius of the
sphere $R$ is large and the angular velocities $\Omega$ (which are chemical potential for 
angular momenta)  are small. In particular, our approximation should capture the term which
encodes the information about the helicity of the flat space thermal vacuum.

We will do this via a hydrodynamic approximation \cite{Bhattacharyya:2007vs,Jensen:2012jh,Banerjee:2012iz} - 
to this end let us begin by rewriting the sphere partition function as 
\begin{equation}
\begin{split}
\mathcal{Z}[\Omega] &\equiv \text{Tr}_{S^{2n-1}\times\mathbb{R}}\ \text{Exp} \brk{ -\frac{\prn{H-\mu Q-\sum_{a=1}^n \Omega_a L_a}}{T}}\\
 &= \text{Tr}\ \text{Exp} \brk{ \frac{\prn{u^\mu P_\mu+\hat{\mu} Q}}{\hat{T}}}\\
\end{split}
\end{equation}
with 
\begin{equation}
\begin{split}
\hat{\mu} &\equiv \gamma \mu \quad ,\qquad \hat{T} \equiv \gamma T\\
u^\mu\partial_\mu &= \gamma \brk{\partial_t+\sum_{a=1}^n\Omega_a
\partial_{\phi_a}}\\
\gamma &\equiv (1-v^2)^{-1/2} \qquad
  v^2 \equiv \sum_{a=1}^n(r_a\Omega_a)^2\\
\end{split}
\end{equation}
This way of writing makes clear that the partition function with $\Omega$s turned on is just a `locally-boosted' version
of the usual partition function where the local-boost is given by $u^\mu$, the local temperature by $\hat{T}$
and the local chemical potential by $\hat{\mu}$. The hydrodynamic approximation consists of writing
\begin{equation}
\begin{split}
\ln\ \mathcal{Z}[\Omega]\quad  &\approx\quad \ln\ \mathcal{Z}_{Hydro}[u^\mu,\hat{T},\hat{\mu}] =-\int_{S^{2n-1}}\frac{1}{\hat{T}}\mathcal{G}^t
\end{split}
\end{equation}
where $\mathcal{G}^\mu$ is the Gibbs free-energy current in hydrodynamics whose t-component is just the Gibbs free-energy density.
In $d=2$, we have 
\begin{equation}
\begin{split}
\ln\ \mathcal{Z}[\Omega]\quad  \approx\ln \mathcal{Z}_{Hydro}
&= -\int_{S^1}\frac{1}{\hat{T}}\mathcal{G}^t \\
\langle H \rangle =  -\int_{S^1} T^t_t \quad &,\ \quad \langle L_a \rangle = \int_{S^1} T^t_{\phi_a}\\
\langle Q \rangle = \int_{S^1} J^t \quad &,\ \quad  S = \int_{S^1} (J_S)^t \\
\end{split}
\end{equation}
with
\begin{equation}
 \begin{split}
  \prn{\mathcal{G}^{\alpha}}_{_{d=2}} &= -\hat{p}u^\alpha + \hat{\mathfrak{F}}^\omega_{anom} V^\alpha \\
 \prn{T^{\alpha\beta}}_{_{d=2}} &= \hat{p}\brk{g^{\alpha\beta}+2u^\alpha u^\beta} - \hat{\mathfrak{F}}^\omega_{anom}(u^\alpha V^\beta+u^\beta V^\alpha) \\
 \prn{J^{\alpha}}_{_{d=2}} &= \frac{\partial\hat{p}}{\partial\hat{\mu}}u^\alpha 
    - \frac{\partial\hat{\mathfrak{F}}^\omega_{anom}}{\partial\hat{\mu}}V^\alpha \\
 \prn{J_S^{\alpha}}_{_{d=2}} &= \frac{\partial\hat{p}}{\partial\hat{T}}u^\alpha 
    - \frac{\partial\hat{\mathfrak{F}}^\omega_{anom}}{\partial\hat{T}}V^\alpha \\
 \end{split}
\end{equation}
where $\hat{p}$ is the local pressure and $V^{\alpha}$ is the Hodge-dual\footnote{Thus, $V^\mu$ is given by 
\begin{equation*}
 \begin{split}
 V&\equiv V_\mu dx^\mu \equiv \bar{u}= \epsilon_{\mu\nu}u^\mu dx^\nu = \gamma R\brk{d\phi-\Omega dt} \\
 V^\alpha \partial_{\alpha} &= \frac{2\pi R^2\gamma\Omega}{\forall_{S^1}} \brk{ \frac{\partial}{\partial t}+\frac{1}{R^2\Omega}\frac{\partial}{\partial \phi}}
  \end{split}
\end{equation*}} of $u^\alpha$ ( we will denote Hodge-duals by an overbar henceforth). 

These expressions make it clear that Cardy approximation can be thought of in hydrodynamic terms.
A more general lesson is that when the sphere is large, the sphere partition function is 
essentially dominated by a rotating fluid configuration. In the rest of this section, we
will use this to actually derive the thermal helicity in higher dimensions in terms of hydrodynamic 
coefficients of the system.

We are interested in fluids living on a sphere $S^{2n-1}$. We 
consider a rigidly rotating velocity configuration for the fluid on the sphere
\begin{equation}
\begin{gathered}
  u^\mu\partial_\mu = \gamma \brk{\partial_t+\sum_{a=1}^n\Omega_a
\partial_{\phi_a}}\\
  u=u_\mu \dr x^\mu = \gamma \brk{-\dr t+\sum_{a=1}^n r_a^2 \Omega_a \dr
\phi_a}\\
  \gamma = (1-v^2)^{-1/2} \qquad
  v^2 = \sum_{a=1}^n(r_a\Omega_a)^2
\end{gathered}
\end{equation}
which is a flow with zero shear and zero expansion. The acceleration and vorticity are given by 
\begin{equation}
\begin{split}
a &= -\gamma^{-1}d\gamma= -\sum_{a=1}^n (\gamma\Omega_a)^2\ r_a dr_a\\
&=  -\sum_{i=1}^{n-1} \gamma^2\prn{\Omega_i^2-\Omega_n^2}\ r_i dr_i \\
\omega &=\gamma^{-1}d\gamma\wedge u +\sum_{a=1}^n \gamma\Omega_a\ r_a dr_a\wedge
d\phi_a = \sum_{a=1}^n \gamma\Omega_a\ r_a dr_a\wedge
\brk{ d\phi_a + \gamma\Omega_a u }\\
\end{split}
\end{equation}
which satisfy $du=2\omega+a\wedge u$. Further, in $d=2n$ spacetime dimensions, let us define a $2n-1$ form $\bar{V}\equiv
(2\omega)^{n-1}\wedge u$. This is the Hodge-dual of a 1-form $V_\mu$ which can be calculated as
\begin{equation}
\begin{split}
V^\mu\partial_\mu &=\frac{\prod_{k=1}^{n}(2\pi R^2 \gamma\Omega_k )}{\forall_{2n-1}}
\left[\partial_t+\sum_{a=1}^n\frac{1}{R^2\Omega_a}\partial_{\phi_a}\right] \\
\end{split}
\end{equation}
It is easily checked that $u_\mu V^\mu =0$.

Following \cite{Loganayagam:2011mu}, we write the Gibbs free energy current in hydrodynamics 
in the form
\begin{equation}
 \begin{split}
  \prn{\mathcal{G}^{\alpha}}_{_{d=2n}} &= -\hat{p}u^\alpha +\ldots+ \hat{\mathfrak{F}}^\omega_{anom} V^\alpha +\ldots \\
 \end{split}
\end{equation}
where we have kept the leading order parity odd and parity even pieces.

The coefficient $\hat{\mathfrak{F}}^\omega_{anom}$ is often termed the chiral vortical coefficient and is 
well-studied in the context of hydrodynamics with anomalies. We will refer the reader to the references 
given in the introduction for various results concerning this coefficient. For our present purposes,
we will only need the following statements : first of all, using general thermodynamic arguments,
one can show\cite{Loganayagam:2011mu,Banerjee:2012cr} that the coefficient $\hat{\mathfrak{F}}^\omega_{anom}$ is
a homogenous $(n+1)$th degree polynomial in temperature and chemical potential. Further, the authors of 
\cite{Loganayagam:2012pz} conjectured that the coefficients appearing in this polynomial are in fact 
anomaly coefficients of the underlying theory, the precise relation being given by a replacement rule
of the form\eqref{eq:anomFP}.

We are now ready to compute the sphere partition function by integrating the zeroeth component of 
Gibbs current over the sphere. This integral can essentially be performed by reducing it to an 
integral of $\gamma^{2n}$ over the sphere and then using the formula\cite{Bhattacharyya:2007vs,2011arXiv1103.5759C}
\begin{equation}\begin{split}\int_{S^{2n-1}} \gamma^{2n} = \frac{\forall_{S^{2n-1}}
}{\prod_{k=1}^{n}(1-R^2\Omega_k^2)}= \frac{2\pi^n R^{2n-1}
}{(n-1)!\prod_{k=1}^{n}(1-R^2\Omega_k^2)} \end{split}\end{equation}
This gives the $\hat{\mathfrak{F}}^\omega_{anom}$ contribution to the sphere partition function as 
\begin{equation}
\begin{split}
\prn{\ln \mathcal{Z}[\Omega]}_{\hat{\mathfrak{F}}^\omega_{anom}} 
&\approx- \frac{1}{T}\left[ \hat{\mathfrak{F}}_{anom}^\omega[T,\mu] \prod_{b=1}^n\frac{2\pi R^2\Omega_b }{1-R^2\Omega_b^2}\right]\\
\end{split}
\end{equation}
where $\{T,\mu\}$ are the global temperature and chemical potential respectively. Further, if the theory under question is 
a CFT, then the pressure contribution can also be integrated to give the expressions in \eqref{eq:higherdim}. We can now 
compute the thermal helicity by differentiating the above contribution with respect to $\Omega$s, setting $\Omega$s
to zero and then taking the flat space limit. This  leads to the result  
\[ \hat{\mathfrak{F}}_{anom}^\omega[T,\mu]=\mathfrak{F}_{anom}^\omega[T,\mu] \]
thus identifying the function appearing in thermal helicity with the chiral vortical coefficient.  
Note that there are no further contributions to the thermal 
helicity from the higher derivative contributions in hydrodynamics since all such contributions
either come with more factors of $\Omega$ or with further factors of $1/R$ and all such contributions vanish after
we take the flat space limit.

We are essentially done  - this relation between thermal helicity and chiral vortical coefficient means that we 
can directly translate the entire theory of chiral vortical coefficient in the hydrodynamics literature to a 
theory of thermal helicity. In particular, this leads to the surprising assertion : thermal helicity is 
forced to always be a polynomial in temperature and chemical potential. Further, if the conjecture made by 
authors of \cite{Loganayagam:2012pz} is correct, then the coefficients appearing in thermal helicity are 
directly related to the anomaly coefficients. In the rest of the paper, we will use this relation between
thermal helicity and chiral vortical coefficient to
test this conjecture against free chiral theories in various dimensions. 

\section{Chiral fermion on $S^{2n-1}\times R$}
\label{sec:anomFermi}

In this section, we are interested in computing the thermal helicity of free fermion theories.
Since the thermal vacuum of free Dirac fermions is invariant under parity we can focus on just
free chiral fermions.

As we had argued in the previous sections, the calculation of 
thermal helicity is same as the calculation of the transport coefficient $\mathfrak{F}^\omega_{anom}$
which for free chiral theories was computed in \cite{Loganayagam:2012pz} via adiabaticity
arguments. According to the authors of \cite{Loganayagam:2012pz}, in $d=2n$ spacetime dimensions
\begin{equation}\label{eq:FWeyl}
\begin{split}
\prn{{\mathfrak{F}}^\omega_{anom}}_{d=2n}
&=- 2\pi\sum_{species}  \chi_{_{d=2n}} \brk{\frac{\frac{\tau}{2}T}{\sin \frac{\tau}{2}T}e^{\frac{\tau}{2\pi}q\mu}}_{\tau^{n+1}} \\
\end{split}
\end{equation}
where the subscript $\tau^{n+1}$ denotes that one needs to Taylor-expand the expression in brackets
near $\tau=0$ and take the coefficient of $\tau^{n+1}$. Every particle/anti-particle pair of Weyl 
fermions contribute a single term in the above sum where  $\{\chi_{_{d=2n}},q\}$ are the chirality 
and charge of the corresponding particle.\footnote{It is easily checked that the expression does not
change if we switch particle with the anti-particle : this corresponds to 
\[ \chi_{_{d=2n}}\mapsto (-1)^{n+1}\chi_{_{d=2n}}\quad ,\quad q \mapsto -q \]. }

There are two remarks about eqn.\eqref{eq:FWeyl} made in \cite{Loganayagam:2012pz} that we will
need in the following. The first is that eqn.\eqref{eq:FWeyl} can be simply obtained from the 
anomaly polynomial of the given free fermion theory via the substitution rule in eqn.\eqref{eq:anomFP}.
The second useful result from \cite{Loganayagam:2012pz} is that  one can split the particle
and the anti-particle contributions in eqn.\eqref{eq:FWeyl} to write
\begin{equation}\label{eq:FWeyl2}
\begin{split}
\prn{{\mathfrak{F}}^\omega_{anom}}_{d=2n} &= -T\sum_{\text{particles}} \frac{ \chi_{_{d=2n}} }{(n-1)!} \int_0^\infty \frac{dE_p }{2\pi}  \prn{\frac{E_p}{2\pi}}^{n-1}\ 
\ln\brk{1+ e^{-\beta\prn{E_p-q\mu}}} \\
\end{split}
\end{equation}
where each particle/anti-particle pair contributes two terms  in the sum above.

The thermal helicity of a free fermion theory can be now computed by substituting \eqref{eq:FWeyl2} into
\eqref{eq:heliF}, 
\begin{equation}
\begin{split}
\frac{1}{\text{Vol}_{2n-1}} &\langle \mathfrak{L}_{12} \mathfrak{L}_{34}\ldots \mathfrak{L}_{2n-3,2n-2} \mathcal{P}_{2n-1} \rangle \\
&=  \sum_{\text{particles}}  \chi_{_{d=2n}}  2^{n-1}\int_0^\infty \frac{dE_p }{2\pi\beta}  \prn{\frac{E_p}{2\pi\beta}}^{n-1}\ 
\ln\brk{1+ e^{-\beta\prn{E_p-q\mu}}} \\
\end{split}
\end{equation}

The aim of this section is to give a microscopic derivation of this formula by a direct computation
of the sphere partition function. Equivalently we want to give a direct microsocopic proof of the results of
\cite{Loganayagam:2012pz}. We are interested in computing 
\[\mathcal{Z} \equiv Tr\ \exp \left[-\beta\left(H-\mu Q-\sum_{i=1}^n\Omega_i
J_i\right)\right] \]
We will do this in the free chiral theories by using state-operator correspondence\cite{Aharony:2003sx}
and counting operators instead.

The essential idea is to count the operators modulo those that vanish because of equations of motion.
For free theories it is convenient to first calculate the contribution in the single particle sector 
and then `multi-particle' it using Fermi-Dirac(or Bose-Einstein) distribution. This last operation 
is often also termed plethystic exponentiation. 

Let us begin by introducing the notation
\begin{equation}
\begin{split}
x &\equiv e^{-R^{-1}\beta} \ ,\qquad  y\equiv e^{\beta\mu}\ \quad\text{and}\quad\  z_i\equiv e^{\beta\Omega_i}=x^{-R\Omega_i}
\end{split}
\end{equation}
We want to first differentiate with respect to $\Omega_i$, then set it to zero followed by taking $R\to\infty$ limit.
In terms of the variables $\{x,y,z_i\}$, this translates to differentiating first with respect to $z_i$ then set  
$z_i\to 1$ followed by the limit $x\to 1$.

We begin by enumerating the letter partition functions for elementary letters and the
equation of motion for a single species of chiral fermion in the table~\S\ref{tab:Weyl}.
\begin{table}[h]
\begin{tabular}[c]{||c|c||c|c||}
\hline
Letter & $\mathcal{Z}_{\text{Letter}}$ & Letter & $\mathcal{Z}_{\text{Letter}}$\\
\hline
&  & &\\
$\partial_i$ & $xz_i $ &$\partial\psi$ & $y^{-q}x^{n+\frac{1}{2}}\Upsilon^F_{(-1)^{n} \chi_{_{d=2n}}}(z)$\\
&  & &\\
$\bar{\partial}_i$ & $xz_i^{-1} $ & $\partial\bar{\psi}$ &  $y^qx^{n+\frac{1}{2}}\Upsilon^F_{- \chi_{_{d=2n}}}(z)$\\
&  & &\\
$\psi$ & $y^{-q}x^{n-\frac{1}{2}}\Upsilon^F_{(-1)^{n+1} \chi_{_{d=2n}}}(z)$ & & \\
&  & &\\
$\bar{\psi}$ &  $y^qx^{n-\frac{1}{2}}\Upsilon^F_{ \chi_{_{d=2n}}}(z)$ & & \\
&  & &\\
\hline 
\end{tabular}
\caption{\label{tab:Weyl} Letter Partition functions for elementary letters and eqns. of motion :
Weyl fermion in $d=2n$ spacetime dimensions.}
\end{table}
In this table  $\{ \chi_{_{d=2n}},q\}$ denote the chirality and charge of the particle respectively while
the symbols $\Upsilon^F_{\pm}(z)$ denote the $so(2n)$ characters of the chiral and the anti-chiral 
fermions.  
\begin{equation}
\begin{split}
\Upsilon^F_{\pm}(z)\equiv
\frac{1}{2}\left[\prod_{i=1}^{n}\left(\sqrt{z_i}+\frac{1}{\sqrt{z_i}}\right)
\pm\prod_{i=1}^{n}\left(\sqrt{z_i}-\frac{1}{\sqrt{z_i}}\right)\right]
\end{split}
\end{equation}
In the table above we have used the fact that the complex conjugate of a chiral fermion in
$2n$ dimensions is chiral if $n$ is odd and antichiral if $n$ is even. Further 
acting by a vector (in this case the derivative operator) flips the chirality of a Weyl spinor.

The single particle partition function follows by looking at words formed by
acting an arbitrary number of derivatives on the letters $\psi,\bar{\psi}$  and then
subtracting out the words formed by the equation of motion
\begin{equation}
\begin{split}
&\mathcal{Z}^{\text{1-particle}}_F \\
&\quad= \sum_{\text{species}}\left[\frac{
y^qx^{n-\frac{1}{2}}\Upsilon^F_{\chi_{_{d=2n}}}(z)+y^{-q}x^{n-\frac{1}{2}}\Upsilon^F_{(-1)^{n+1}\chi_{_{d=2n}}}(z)}
{\prod_{i=1}^{n}(1-xz_i)(1-\frac{x}{z_i})}\right.\\
&\qquad \left.-\frac{y^qx^{n+\frac{1}{2}}\Upsilon^F_{-\chi_{_{d=2n}}}(z)+y^{-q}x^{n+\frac{1}{2}}\Upsilon^F_{(-1)^{n
}\chi_{_{d=2n}}}(z)}{\prod_{i=1}^{n}(1-xz_i)(1-\frac{x}{z_i})}\right]\\
&\quad = \sum_{\text{particles}}y^qx^{n-\frac{1}{2}}\left[\frac{
\Upsilon^F_{\chi_{_{d=2n}}}-x\Upsilon^F_{-\chi_{_{d=2n}}}} {\prod_{i=1}^{n}(1-xz_i)(1-\frac{x}{z_i})}\right]\\
&\quad = \frac{1}{2}x^{n-\frac{1}{2}}(1-x)
\prod_{i=1}^{n}\frac{\left(\sqrt{z_i}+\frac{1}{\sqrt{z_i}
}\right)}{(1-xz_i)(1-\frac{x}{z_i})}\sum_{\text{particles}}y^q\\
&\qquad+\frac{1}{2}x^{n-\frac{1}{2}}(1+x) \prod_{i=1}^{n}\frac{\left(\sqrt{z_i}-\frac{1}{\sqrt{z_i}}\right)}{
(1-xz_i)(1-\frac{x}{z_i})} \sum_{\text{particles}}\chi_{_{d=2n}} y^q\\
\end{split}
\end{equation}
Before proceeding note that the single particle partition function naturally splits into two parts -
as is easily checked the term  proportional to $\sum_{\text{particles}}y^q$ is half the contribution
coming from a Dirac fermion. The presence of the second term makes it clear that the free energy of the
ideal Weyl gas is not just half the free-energy of the ideal Dirac gas as one might have naively expected. 
This fact (which is equivalent to the statement that the Weyl path-integral is
not the square-root of Dirac path-integral) is closely related to the underlying
anomalies in the theory with just a single species of Weyl fermion. The effect of these anomalies
at the level of free-energy is captured by the extra contribution proportional to
$\sum_{\text{particles}}\chi_{_{d=2n}} y^q$. 

The sphere partition function of a free Weyl gas is now given by multi-particling the
above single particle partition function fermionically
\begin{equation}
\begin{split}
\ln\ \mathcal{Z}[\Omega] &= \sum_{j=1}^{\infty}\frac{(-1)^{j-1}}{j}
\mathcal{Z}^{\text{1-particle}}_F(x^j,y^j,(z_i)^j) \\
\end{split}
\end{equation}
and we are interested in calculating the thermal helicity via 
\begin{equation}
\begin{split}
\frac{1}{\text{Vol}_{2n-1}} 
&\langle \mathfrak{L}_{12} \mathfrak{L}_{34}\ldots \mathfrak{L}_{2n-3,2n-2} \mathcal{P}_{2n-1} \rangle\\
&= \lim_{R\to \infty} \frac{T^n}{R\forall_{2n-1}}\brk{\prn{\prod_{a=1}^n\frac{\partial}{\partial \Omega_a}}
\ln \mathcal{Z}[\Omega] }_{\Omega=0} \\
\end{split}
\end{equation}
This evaluates to
\begin{equation}
\begin{split}
\frac{1}{\text{Vol}_{2n-1}} 
&\langle \mathfrak{L}_{12} \mathfrak{L}_{34}\ldots \mathfrak{L}_{2n-3,2n-2} \mathcal{P}_{2n-1} \rangle\\
&= - \sum_{\text{particles}}\sum_{j=1}^{\infty} \chi_{_{d=2n}}j^{n-1} (-y^q)^j
\brk{\lim_{R\to \infty} \frac{1}{R\forall_{2n-1}}\frac{x^{j(n-\frac{1}{2})}(1+x^j) }{
2(1-x^j)^{2n}} }\\
&= -\sum_{\text{particles}}\chi_{_{d=2n}}\frac{(n-1)!}{2\pi^n\beta^{2n}}
\sum_{j=1}^{\infty}  \frac{ (-y^q)^j }{ j^{n+1}} \\
\end{split}
\end{equation}
where we  have used $x\approx 1-R^{-1}\beta$ for large $R$. The infinite sum over $j$ can be 
converted into an integral
\begin{equation}
\begin{split}
\frac{1}{\text{Vol}_{2n-1}} 
&\langle \mathfrak{L}_{12} \mathfrak{L}_{34}\ldots \mathfrak{L}_{2n-3,2n-2} \mathcal{P}_{2n-1} \rangle\\
&= -\sum_{\text{particles}}\chi_{_{d=2n}} 2^{n-1}
\sum_{j=1}^{\infty}   \frac{1}{j} \int_0^\infty \frac{dE_p}{2\pi\beta} \brk{-y^q e^{-\beta E_p}}^j \prn{\frac{E_p}{2\pi\beta}}^{n-1} \\
&=  \sum_{\text{particles}}  \chi_{_{d=2n}} 2^{n-1} \int_0^\infty \frac{dE_p }{2\pi\beta}  \prn{\frac{E_p}{2\pi\beta}}^{n-1}\ 
\ln\brk{1+ e^{-\beta\prn{E_p-q\mu}}} \\
\end{split}
\end{equation}
which is the result we wanted to prove. This completes the direct microscopic derivation of the results
of \cite{Loganayagam:2012pz}.


\section{Chiral Boson on $S^{2n-1}\times R$ (with $n$ odd)}
\label{sec:anomBose}

In this section, we will compute the thermal helicity of a free theory of chiral bosons (i.e.,
abelian self-dual or anti-self dual forms) . Since chiral bosons exist only when $n$ is
odd (where $\ast^2=-(-1)^{p(d-p)}=1$ acting on $p=n$-forms in $d=2n$ spacetime dimensions),
we will assume $n$ to be odd in the rest of this section. Since these bosons are their own
anti-particles we will not distinguish between a sum over species vs. sum over particles.

Before proceeding to the calculation, let us try to see what the answer should be if the 
substitution rule \eqref{eq:anomFP} holds for chiral bosons. The anomaly polynomial of
chiral bosons is given in terms of the Hirzebruch genus (or L-genus) as we review in 
the appendix~\ref{app:anom} on anomaly polynomials. Applying the substitution rule
on this anomaly polynomial we get
\begin{equation}\label{eq:FBose}
\begin{split}
\prn{{\mathfrak{F}}^\omega_{anom}}_{d=2n}
&= 2\pi\sum_{particles}  \chi_{_{d=2n}}\frac{1}{8} \brk{\frac{\tau T}{\tan \tau T}}_{\tau^{n+1}} \\
&= -T\sum_{\text{particles}} \frac{2^{n-1} \chi_{_{d=2n}} }{(n-1)!} \int_0^\infty \frac{dE_p }{2\pi}  \prn{\frac{E_p}{2\pi}}^{n-1}\ 
\ln\brk{1- e^{-\beta E_p}}^{-1} \\
\end{split}
\end{equation}
where the subscript $\tau^{n+1}$ denotes that one has to pick the coefficient of $\tau^{n+1}$ in the Taylor
expansion in $\tau$. In the second line we have rewritten the expected answer in terms of a Bose-Einstein
integral using identities derived in appendix~\ref{app:BoseFermiMoments}.

This is equivalent to the statement that the thermal helicity of a theory of chiral bosons is given by 
\begin{equation}
\begin{split}
\frac{1}{\text{Vol}_{2n-1}} &\langle \mathfrak{L}_{12} \mathfrak{L}_{34}\ldots \mathfrak{L}_{2n-3,2n-2} \mathcal{P}_{2n-1} \rangle \\
&=  \sum_{\text{particles}}  \chi_{_{d=2n}}  2^{2n-2}\int_0^\infty \frac{dE_p }{2\pi\beta}  \prn{\frac{E_p}{2\pi\beta}}^{n-1}\ 
\ln\brk{1- e^{-\beta E_p}}^{-1} \\
\end{split}
\end{equation}
Our aim is to derive this formula using sphere partition function. As before, 
let $x\equiv e^{-R^{-1}\beta}$ and $z_i\equiv e^{\beta\Omega_i}=x^{-R\Omega_i}$.
We are interested in computing 
\[\mathcal{Z} \equiv Tr\ \exp \left[-\beta\left(H-\sum_{i=1}^n\Omega_i
J_i\right)\right] \]
for chiral bosons in $2n$ spacetime dimensions. 

The trace can again be performed by using state-operator correspondence
\cite{Aharony:2003sx} and counting operators instead. The relevant 
letter partition functions for elementary letters and the equation of
motion are given in table~\S\ref{tab:Bose}
\begin{table}[h]
\begin{tabular}[c]{||c|c||}
\hline
Letter & $\mathcal{Z}_{\text{Letter}}$\\
\hline
& \\
$\partial_i$ & $xz_i $ \\
& \\
$\bar{\partial}_i$ & $xz_i^{-1} $ \\
& \\
$F=\pm \ast F$ & $x^{n}\Upsilon^B_{n\pm}(z)$\\
& \\
\hline
& \\
$d^kF$ & $x^{n+k}\Upsilon^B_{n+k}(z)=x^{n+k}\Upsilon^B_{n-k}(z)$\\
& \\
\hline
\end{tabular}
\caption{\label{tab:Bose} Letter Partition functions for elementary letters and eqns. of motion :
Chiral p-forms in $d=2n$ spacetime dimensions (with $n$ odd).}
\end{table}
In the table,  $\Upsilon^B_{k}(z)$ are the $so(2n)$ characters of the $k$-forms for 
$k\neq n$. Note that Hodge duality implies that the character of $k$-form is equal 
to the character of $n-k$ form. 
\[ \Upsilon^B_{2n-k}(z)=\Upsilon^B_{k}(z). \]
$\Upsilon^B_{n\pm}(z)$ are the $so(2n)$ characters of the chiral and antichiral
$n$-forms respectively. All these characters are completely determined by 
their dominant weights. For k-forms, the dominant weights are given by
\[\prn{\lambda_1,\lambda_2,\ldots,\lambda_n}_k=(\underbrace{1,1,\ldots}_{\text{k
times}}, \underbrace{0 ,0, \ldots}_{\text{(n-k) times}} ) \]
For chiral and antichiral $n$-forms, the dominant weights are
\[\prn{\lambda_1,\lambda_2,\ldots,\lambda_n}_{n\pm}=(\underbrace{1,1,\ldots}_{
\text{(n-1) times}},\pm 1) \]
The corresponding characters are given by the Weyl character formula for
$so(2n)$
\[ \Upsilon^B_{\{\lambda_i\}} \equiv 
\frac{
|z_j^{\lambda_i+n-i}+z_j^{-\prn{\lambda_i+n-i}}|
+|z_j^{\lambda_i+n-i}-z_j^{-\prn{\lambda_i+n-i}}|}
{|z_j^{n-i}+z_j^{-\prn{n-i}}|}\]
where $|A_{ij}|$ denotes the determinant of the matrix formed with numbers
$A_{ij}$ in its $(i,j)^{th}$ position. The second term in the numerator 
vanishes for all $k$-forms with $k\neq n$. For the $n$-forms we have 
\[\brk{\frac{
|z_j^{\lambda_i+n-i}-z_j^{-\prn{\lambda_i+n-i}}|}
{|z_j^{n-i}+z_j^{-\prn{n-i}}|}}_{n\pm}=\pm \frac{
|z_j^{n+1-i}-z_j^{-\prn{n+1-i}}|}
{|z_j^{n-i}+z_j^{-\prn{n-i}}|} = \pm\frac{1}{2}
\prod_{i=1}^n\prn{z_i-z_i^{-1}}\]

It is sometimes more convenient to work with the following generating function for 
the characters of the $k$-form representations.
\begin{equation}
\begin{split}
\prod_{i=1}^n(1+tz_i)(1+tz_i^{-1})&= \sum_{j=0}^{2n}t^j \Upsilon^B_{j}(z) \\
&=
t^n\brk{\Upsilon^B_{n+}(z)+\Upsilon^B_{n-}(z)+\sum_{k=1}^n(t^k+t^{-k})\Upsilon^B_{n-k}(z)}\\
\prod_{i=1}^n\prn{z_i-z_i^{-1}}&= \Upsilon^B_{n+}(z)-\Upsilon^B_{n-}(z)
\end{split}
\end{equation}
These two formulae are sufficient to derive all the characters that we need.

The single particle partition function follows by looking at words formed by
acting an arbitrary number of derivatives on the letter $F$  and then subtracting
out the words formed by the equation of motion. For the chiral bosons
the equations of motion are $dF=0$, but all these equations are not independent.
They obey a constraint $d^2F=0$ and these constraints are further constrained by
$d^3F=0$ and so on. So,the final letter partition function can be written in
terms of an alternating sum
\begin{equation}
\begin{split}
\mathcal{Z}^{\text{1-particle}}_{B} &= \sum_{\text{particles}}
 \frac{
x^n\Upsilon^B_{n,\chi_{_{d=2n}}}(z)-x^{n+1}\Upsilon^B_{n-1}(z)+x^{n+2}\Upsilon^B_{n-2}(z)\ldots+(-1)^n
x^{2n}\Upsilon^B_{0}(z)
}{\prod_{i=1}^{n}(1-xz_i)(1-xz_i^{-1})}\\
&=  \sum_{\text{particles}}\frac{x^n}{\prod_{i=1}^{n}(1-xz_i)(1-xz_i^{-1})}\times\prn{\frac{1}{2}\Upsilon^B_{n+}(z)+\frac{1}{2}\Upsilon^B_{n-}(z)+\sum_{k=1}^n(-x)^k\Upsilon^B_{n-k}
(z)}\\
&\qquad+\frac{1}{2}\frac{x^n}{\prod_{i=1}^{
n}(1-xz_i)(1-xz_i^{-1})} \sum_{\text{particles}}\chi_{_{d=2n}}\brk{\Upsilon^B_{n+}(z)-\Upsilon^B_{n-}(z)}\\
&=  \sum_{\text{particles}}\frac{x^n}{\prod_{i=1}^{n}(1-xz_i)(1-xz_i^{-1})}\times\prn{\frac{1}{2}\Upsilon^B_{n+}(z)+\frac{1}{2}\Upsilon^B_{n-}(z)+\sum_{k=1}^n(-x)^k\Upsilon^B_{n-k}
(z)}\\
&\qquad+\frac{1}{2}\sum_{\text{particles}}\chi_{_{d=2n}}x^n\prod_{i=1}^{n}\frac{(z_i-z_i^{-1})}{(1-xz_i)(1-xz_i^{-1})}
\end{split}
\end{equation}
Compare this expression with the conformal character of the corresponding short
representation given by putting $l=1$ in equation(3.26) of \cite{Dolan:2005wy}. 
Like the computation in the previous section, the single particle partition function
splits into a (parity even) non-chiral part and a (parity odd) chiral part.

The sphere partition function of a free chiral boson gas is now given by bosonically 
multi-particling this single particle partition function 
\begin{equation}
\begin{split}
\ln\ \mathcal{Z}[\Omega] &= \sum_{j=1}^{\infty}\frac{1}{j}
\mathcal{Z}^{\text{1-particle}}_B(x^j,y^j,(z_i)^j) \\
\end{split}
\end{equation}
and we are interested in calculating the thermal helicity via 
\begin{equation}
\begin{split}
\frac{1}{\text{Vol}_{2n-1}} 
&\langle \mathfrak{L}_{12} \mathfrak{L}_{34}\ldots \mathfrak{L}_{2n-3,2n-2} \mathcal{P}_{2n-1} \rangle\\
&= \lim_{R\to \infty} \frac{T^n}{R\forall_{2n-1}}\brk{\prn{\prod_{a=1}^n\frac{\partial}{\partial \Omega_a}}
\ln \mathcal{Z}[\Omega] }_{\Omega=0} \\
\end{split}
\end{equation}
This evaluates to
\begin{equation}
\begin{split}
\frac{1}{\text{Vol}_{2n-1}} 
&\langle \mathfrak{L}_{12} \mathfrak{L}_{34}\ldots \mathfrak{L}_{2n-3,2n-2} \mathcal{P}_{2n-1} \rangle\\
&=  \sum_{\text{particles}}\sum_{j=1}^{\infty} \chi_{_{d=2n}}2^{n-1} j^{n-1} 
\brk{\lim_{R\to \infty} \frac{1}{R\forall_{2n-1}}\frac{x^{j n} }{
(1-x^j)^{2n}} }\\
&=  \sum_{\text{particles}}  \chi_{_{d=2n}} 2^{2n-2} \int_0^\infty \frac{dE_p }{2\pi\beta}  \prn{\frac{E_p}{2\pi\beta}}^{n-1}\ 
\ln\brk{1- e^{-\beta E_p}}^{-1} \\
\end{split}
\end{equation}
which is the result we wanted to prove.

Note that this section along with the previous section prove our statements about thermal helicity
for arbitrary even dimensional free theories. We find that the thermal helicity gets contributions
only from chiral fields and that this contribution is a polynomial in $T$ and $\mu$ with 
coefficients drawn from the anomaly polynomial of the theory. This is in accordance 
with  the conjecture \eqref{eq:anomFP} made in \cite{Loganayagam:2012pz} and is
consistent with the recent proof of this conjecture in $d=2$ and $d=4$
by the authors of \cite{Jensen:2012kj}. 

\section{Chiral gravitino on $S^{2n-1}\times R$ }
\label{sec:anomGravitino}
In this section we will begin by extending our sphere partition function calculations to the 
case of Chiral gravitino. In doing so, we will find that it violates the relation \eqref{eq:anomFP}.
We will then give an alternate derivation of the same result using the adiabaticity methods
introduced in \cite{Loganayagam:2012pz}.

We will begin by computing the letter partition functions for elementary letters and the equation of
motion. The Weyl gravitino is represented by a spin-$\frac{3}{2}$ Rarita-Schwinger field $\psi_\mu$
which is a Weyl spinor valued 1-form with dimension $n-\frac{1}{2}$. But, $\psi_\mu$ itself 
is not gauge-invariant, rather it transforms as $\psi_\mu\mapsto \psi_\mu+\partial_\mu \vartheta$
where $\vartheta$ is a Weyl spinor field. Since we are interested in counting only the
gauge-invariant operators we will begin by defining 
\[ \Psi_{\mu\nu} \equiv \partial_\mu \psi_\nu - \partial_\nu \psi_\mu \]
which is a gauge invariant Weyl spinor-valued 2-form .

From its definition, it is clear that $ \Psi_{\mu\nu}$ obeys a hierarchy  of 
equations of the type 
\[d^k\Psi=\partial^k_{[\mu_1\mu_2\ldots\mu_k}\Psi_{\mu_{k+1}\mu_{k+2}]}=0\]
very similar to the form fields of the last section. In addition, there is
the equation of motion which is just the Rarita-Schwinger equation  
\[\Gamma^{\mu\nu\lambda} \Psi_{\nu\lambda} =0\]
The Rarita Schwinger equations are all not independent though - they in 
turn obey $\Gamma^{\mu\nu\lambda}\partial_\mu \Psi_{\nu\lambda} =0$.
We enumerate the letter partition functions associated with all these
gauge-invariant operators in table~\S\ref{tab:grav}.

\begin{table}[h]
\begin{tabular}[c]{||c|c||c|c||}
\hline
Letter & $\mathcal{Z}_{\text{Letter}}$ & Letter & $\mathcal{Z}_{\text{Letter}}$\\
\hline
& & & \\
$\partial_i$ & $xz_i $ & $\bar{\partial}_i$ & $xz_i^{-1}$ \\
& & & \\
$\bar{\Psi}_{\mu\nu}$ & $x^{n+\frac{1}{2}}y^q\Upsilon^F_{\chi_{_{d=2n}}}(z)\Upsilon^B_2(z)$ &
$\Psi_{\mu\nu}$ &
$x^{n+\frac{1}{2}}y^{-q}\Upsilon^F_{(-1)^{n+1}\chi_{_{d=2n}}}(z)\Upsilon^B_2(z)$\\
& & & \\ 
\hline
& & & \\
$d^{k-2}\bar{\Psi}$ & $x^{n-\frac{3}{2}+k}y^q\Upsilon^F_{\chi_{_{d=2n}}}(z)\Upsilon^B_{k}(z)$ &
$d^{k-2}\Psi$ &
$x^{n-\frac{3}{2}+k}y^{-q}\Upsilon^F_{(-1)^{n+1}\chi_{_{d=2n}}}(z)\Upsilon^B_{k}(z)$\\
& & & \\ 
\hline 
& & & \\
$ \bar{\Psi}_{\nu\lambda}\Gamma^{\mu\nu\lambda}$ &
$x^{n+\frac{1}{2}}y^q\Upsilon^F_{-\chi_{_{d=2n}}}(z)\Upsilon^B_1(z)$ &  
$\Gamma^{\mu\nu\lambda} \Psi_{\nu\lambda}$ &
$x^{n+\frac{1}{2}}y^{-q}\Upsilon^F_{(-1)^{n}\chi_{_{d=2n}}}(z)\Upsilon^B_1(z)$\\
& & & \\ 
\hline 
& & & \\ 
$ \partial_\mu\bar{\Psi}_{\nu\lambda}\Gamma^{\mu\nu\lambda}$ &
$x^{n+\frac{3}{2}}y^q\Upsilon^F_{-\chi_{_{d=2n}}}(z)$ &  
$\Gamma^{\mu\nu\lambda} \partial_\mu\Psi_{\nu\lambda}$ &
$x^{n+\frac{3}{2}}y^{-q}\Upsilon^F_{(-1)^{n}\chi_{_{d=2n}}}(z)$\\
& & & \\ 
\hline
\end{tabular}
\caption{\label{tab:grav} Letter Partition functions for elementary letters and eqns. of motion :
Chiral gravitino in $d=2n$ spacetime dimensions.}
\end{table}
The notations are as before : $x\equiv e^{-R^{-1}\beta}$, $y\equiv e^{\beta\mu}$, $z_i\equiv e^{\beta\Omega_i}$ and 
$q$ is the charge of the gravitino. $\Upsilon^F_{\pm}(z),\Upsilon^B_k(z)$ are the $so(2n)$ characters of 
the chiral/anti-chiral fermions and  $k$-form respectively. They are given by 
\begin{equation}
\begin{split}
\Upsilon^F_{\pm}(z)&\equiv
\frac{1}{2}\left[\prod_{i=1}^{n}\left(\sqrt{z_i}+\frac{1}{\sqrt{z_i}}\right)
\pm\prod_{i=1}^{n}\left(\sqrt{z_i}-\frac{1}{\sqrt{z_i}}\right)\right]\\
\sum_{j=0}^{2n}t^j \Upsilon^B_{j}(z)&= \prod_{i=1}^n(1+tz_i)(1+tz_i^{-1})  \\
\end{split}
\end{equation}
Further we have used the fact that the complex conjugate of a chiral fermion in
$2n$ dimensions is chiral if $n$ is odd and antichiral if $n$ is even. Acting 
by a odd number of gamma matrices flips the chirality of a Weyl spinor.

The single particle partition function follows by looking at words formed by
acting an arbitrary number of derivatives on the letter $\Psi$  and then subtracting
out the words formed by the equation of motion. They obey $d\Psi=0$, but all
these equations are not independent : there is a constraint $d^2\Psi=0$ and 
these constraints are further constrained by $d^3\Psi=0$ and so on. So,the 
final letter partition function can be written in terms of an alternating sum
\begin{equation}
\begin{split}
\mathcal{Z}^{\text{1-particle}}_{\frac{3}{2}} &=
 \sum_{\text{species}}\prn{y^q\Upsilon^F_{\chi_{_{d=2n}}}+y^{-q}\Upsilon^F_{(-1)^{n+1}\chi_{_{d=2n}}}}\\
&\times\brk{\frac{
x^{n-\frac{3}{2}}\brk{
x^2\Upsilon^B_{2}(z)-x^3 \Upsilon^B_{3}(z)+\ldots + x^{2n}\Upsilon^B_{2n}(z) }
}{\prod_{i=1}^{n}(1-xz_i)(1-xz_i^{-1})}}_{\Psi,d^{k}\Psi}\\
&\quad- \sum_{\text{species}}\prn{y^q\Upsilon^F_{-\chi_{_{d=2n}}}+y^{-q}\Upsilon^F_{(-1)^n\chi_{_{d=2n}}}}\brk{\frac{
x^{n+\frac{1}{2}}\Upsilon^B_{1}(z)-x^{n+\frac{3}{2}}
}{\prod_{i=1}^{n}(1-xz_i)(1-xz_i^{-1})}}_{\Gamma\Psi,\Gamma\partial\Psi}\\
 &=
 \sum_{\text{particles}}x^{n-\frac{3}{2}}y^q\Upsilon^F_{\chi_{_{d=2n}}}\frac{
\brk{
x^2\Upsilon^B_{2}(z)-x^3 \Upsilon^B_{3}(z)+\ldots + x^{2n}\Upsilon^B_{2n}(z) }
}{\prod_{i=1}^{n}(1-xz_i)(1-xz_i^{-1})}\\
&\quad- \sum_{\text{particles}}x^{n+\frac{1}{2}}y^q\Upsilon^F_{-\chi_{_{d=2n}}}\frac{
\Upsilon^B_{1}(z)-x}{\prod_{i=1}^{n}(1-xz_i)(1-xz_i^{-1})}\\
\end{split}
\end{equation}
where the first line counts the operators formed by acting derivatives on $\Psi$ (and
$\bar{\Psi}$ ) mindful of the fact that  $d^k\Psi=0$ and the second line subtracts the
operators that are zero because of Rarita Schwinger equation. Now we use
\[ x^2\Upsilon^B_{2}(z)-x^3 \Upsilon^B_{3}(z)+\ldots + x^{2n}\Upsilon^B_{2n}(z) =
\prod_{i=1}^{n}(1-xz_i)(1-xz_i^{-1})- \brk{
1-x\Upsilon^B_{1}(z)} \]
to write
\begin{equation}
\begin{split}
\mathcal{Z}^{\text{1-particle}}_{\frac{3}{2}} &=
\sum_{\text{particles}}x^{n-\frac{3}{2}} y^q\Upsilon^F_{\chi_{_{d=2n}}}
\brk{1 + \frac{ x\Upsilon^B_{1}(z)-1} {\prod_{i=1}^{n}(1-xz_i)(1-xz_i^{-1})} }\\
&\quad - \sum_{\text{particles}} x^{n+\frac{1}{2}}y^q\Upsilon^F_{-\chi_{_{d=2n}}}
\frac{ \Upsilon^B_{1}(z)-x }{\prod_{i=1}^{n}(1-xz_i)(1-xz_i^{-1})}\\
&=
\frac{1}{2} \sum_{\text{particles}} x^{n-\frac{3}{2}}y^q
\brk{1+\frac{ (1-x)\brk{x\Upsilon^B_{1}(z)-(1+x+x^2)} } {\prod_{i=1}^{n}(1-xz_i)(1-xz_i^{-1})}}
\prod_{i=1}^n\prn{\sqrt{z_i}+\frac{1}{\sqrt{z_i}}}\\
&\quad +\frac{1}{2} \sum_{\text{particles}}\chi_{_{d=2n}} x^{n-\frac{3}{2}}y^q
\brk{1+\frac{ (1+x)\brk{x\Upsilon^B_{1}(z)-(1-x+x^2)} } {\prod_{i=1}^{n}(1-xz_i)(1-xz_i^{-1})}}
\prod_{i=1}^n\prn{\sqrt{z_i}-\frac{1}{\sqrt{z_i}}}\\
\end{split}
\end{equation}
where $\Upsilon^B_{1}(z) = \sum_{i=1}^n \prn{z_i+z_i^{-1}}$ is the $so(2n)$
character of a vector representation. Note that the $\mathcal{Z}^{\text{1-particle}}$
decomposes into a parity-even non-chiral part and a parity odd chiral-part as in the
the case of chiral fermions/chiral forms in previous sections.

We will now multi-particle the above single particle partition function fermionically
\begin{equation}
\begin{split}
\ln\ \mathcal{Z}[\Omega] &= \sum_{j=1}^{\infty}\frac{(-1)^{j-1}}{j}
\mathcal{Z}^{\text{1-particle}}_{\frac{3}{2}}(x^j,y^j,(z_i)^j) \\
\end{split}
\end{equation}
and the corresponding thermal helicity is
\begin{equation}
\begin{split}
\frac{1}{\text{Vol}_{2n-1}} 
&\langle \mathfrak{L}_{12} \mathfrak{L}_{34}\ldots \mathfrak{L}_{2n-3,2n-2} \mathcal{P}_{2n-1} \rangle\\
&= - \sum_{\text{particles}}\sum_{j=1}^{\infty} \chi_{_{d=2n}}j^{n-1} (-y^q)^j \\
&\quad\times \lim_{R\to \infty} \frac{1}{R\forall_{2n-1}}
\frac{1}{2}x^{j(n-\frac{3}{2})} \brk{ 1+ \frac{ (1+x^j)\brk{ 2n x-(1-x+x^2)} }{(1-x^j)^{2n}} } \\
&= -\sum_{\text{particles}}(2n-1)\chi_{_{d=2n}}\frac{(n-1)!}{2\pi^n\beta^{2n}}
\sum_{j=1}^{\infty}  \frac{ (-y^q)^j }{ j^{n+1}} \\
\end{split}
\end{equation}
where we  have used $x\approx 1-R^{-1}\beta$ for large $R$. The infinite sum over $j$ can be 
converted into a Fermi-Dirac integral
\begin{equation}
\begin{split}
\frac{1}{\text{Vol}_{2n-1}} 
&\langle \mathfrak{L}_{12} \mathfrak{L}_{34}\ldots \mathfrak{L}_{2n-3,2n-2} \mathcal{P}_{2n-1} \rangle\\
&= -\sum_{\text{particles}}\chi_{_{d=2n}} (2n-1)2^{n-1}
\sum_{j=1}^{\infty}   \frac{1}{j} \int_0^\infty \frac{dE_p}{2\pi\beta} \brk{-y^q e^{-\beta E_p}}^j \prn{\frac{E_p}{2\pi\beta}}^{n-1} \\
&=  \sum_{\text{particles}}  \chi_{_{d=2n}} (2n-1)2^{n-1} \int_0^\infty \frac{dE_p }{2\pi\beta}  \prn{\frac{E_p}{2\pi\beta}}^{n-1}\ 
\ln\brk{1+ e^{-\beta\prn{E_p-q\mu}}} \\
\end{split}
\end{equation}
Thus we arrive at the result that the thermal helicity of a Weyl gravitino is $(2n-1)$ times the thermal
helicity of a Weyl fermion with the same charge and chirality. This result directly violates the rule 
\eqref{eq:heliP} since the anomaly polynomial of a Weyl gravitino is not proportional to that of a 
Weyl fermion.

Let us now give an alternate (and a quicker) derivation of this result for gravitino using the microscopic
adiabaticity arguments of \cite{Loganayagam:2012pz}. The essential idea behind \cite{Loganayagam:2012pz}
was to find a chiral spectral current that obeys a kind of Liouville theorem (which is crucial for 
second law of thermodynamics to hold). Authors of \cite{Loganayagam:2012pz} concluded that for a fermion
the only combination of magentic field $B$ and vorticity $\omega$ that can appear in the chiral density of states
that is consistent with this adiabaticity argument is of the form
\begin{equation}
\begin{split}
\mathcal{J}_q \sim \chi_{_{d=2n}}\prn{\frac{q B +2\omega E_p}{2\pi}}^{n-1}
\end{split}
\end{equation}
where the prefactor is completely determined by the pure $U(1)^{n+1}$ anomaly contribution of that fermion.
Note that this argument does not rely on the spin of the particle under the question (except for the assumption
that it is a massless chiral fermion ) and hence applies equally well to the case of chiral gravitino.

The next step in the argument is to realise that in $d=2n$ spacetime dimensions the pure $U(1)^{n+1}$ 
anomaly contribution  of a charged chiral gravitino is $(2n-1)$times that of a spin-$\frac{1}{2}$ particle
of same charge and chirality. This along with the adiabaticity constraint  derived by \cite{Loganayagam:2012pz}
implies that all the chiral parts of thermodynamic quantities for a charged chiral gravitino are
$(2n-1)$times that of a spin-$\frac{1}{2}$ particle of same charge and chirality. This extends to 
the transport coefficient $\mathfrak{F}^\omega_{anom}$ and hence to thermal helicity. This
gives an alternate derivation for the result that we obtained above by an explicit counting of states.

\section{Conclusion}\label{sec:conclusion}
In this paper, we have studied various interesting properties of a thermal observeable called 
the thermal helicity. It is defined as the following thermal expectation value 
$< \mathfrak{L}_{12} \mathfrak{L}_{34} ... \mathfrak{L}_{d-3,d-2} \mathcal{P}_{d-1} >$  where
$\mathfrak{L}_{ij}$ denote the angular
momenta, and $\mathcal{P}_i$ denotes the spatial momentum in $d$ spacetime dimensions
with $d$ even. 

Using a variety of arguments, we have shown that the polynomial structure of anomalous transport 
coefficients discovered in \cite{Loganayagam:2011mu,Banerjee:2012cr} has a surprising implication for 
the thermal helicity : thermal helicity per unit volume is essentially forced to be a 
homogeneous polynomial in $T$ and $\mu$. Further, say we assume the conjecture by the authors 
of \cite{Loganayagam:2012pz} relating this polynomial structure to the corresponding
anomaly polynomial  via a replacement rule
\begin{equation}
\begin{split}
\mathfrak{F}_{anom}^\omega[T,\mu] = \mathcal{P}_{anom} \brk{ F \mapsto \mu, p_1(\mathfrak{R}) \mapsto - T^2 , p_{k>1}(\mathfrak{R}) \mapsto 0 }
\end{split}
\end{equation}
Then it follows that the coefficients appearing in the thermal helicity also follow a replacement 
rule given by \eqref{eq:heliP}. By an explicit computation of the sphere partition function and hence the
thermal helicity, we have proved that this conjecture is indeed true in  an arbitrary free theory with chiral
fermions and chiral p-form fields. This statement combined with various AdS/CFT results lead us 
to expect that this rule should be true in general.

Surprisingly, however, by directly computing the thermal helicity, we have shown that a theory of chiral gravitini
violates this replacement rule. This suggests that any general  proof of this rule
should incorporate the assumption that there are no massless higher spin modes in the theory 
under consideration. It would be interesting to examine in detail how the recent proof\cite{Jensen:2012kj}
of the replacement rule in $d=2$ and $d=4$ spacetime dimensions allows for such an exception. 

In $d=2$, since the replacement rule follows from Cardy formula, the violation of the replacement
rule is intimately linked with the violation of the Cardy formula by the theory of free chiral gravitino.
This suggests that what is at play is some deeper principle like modular invariance and it would 
be interesting to know whether this statement can be made more precise. This is especially
intriguing since higher dimensional analogues of modular invariance are ill-understood and 
it is worth exploring whether the deficit in thermal helicity is a useful probe for failure of 
modular invariance.

In the context of AdS/CFT, we have already remarked on the fact that all the existing
computations of the chiral vortical coefficient in AdS$_3$ and AdS$_5$ can be
reformulated as a computation of thermal helicity. In fact, thermal helicity can be
computed more directly from the thermodynamics of AdS Kerr Blackholes in the
presence of various Chern-Simons terms\footnote{See \cite{Tachikawa:2006sz,Bonora:2011gz} 
for a discussion of how the Blackhole thermodyanmics gets modified in the 
presence of Chern-Simons terms.}. It would be interesting to see whether this 
fact can be used to give a general proof that thermal helicity always obeys the 
replacement rule \eqref{eq:heliP}.

Finally as we have emphasised in the introduction, one would hope that 
the results of this  paper serve as a starting point for  generalising 
the analysis of anomalies to non-equilibrium phenomena. Can one
study how thermal helicity behaves away from equilibrium ? Are there 
universal non-equibrium statements that can be made about thermal helicity by linking it 
with anomalies ? We leave such questions to future work.

\subsection*{Acknowledgements}
\label{s:acks}
 It is a pleasure to thank Tatzuo Azeyanagi, Koushik Balasubramanian, Jyotirmoy Bhattacharya,
 Sayantani Bhattacharyya, Jacob L. Bourjaily, Alejandra Castro, Clay Cordova, Stephane  Detournay,
 Nabil Iqbal, Kristan Jensen, Vijay Kumar, Hong Liu, John McGreevy, Shiraz Minwalla, Gim-Seng Ng,
 Suvrat Raju,  Mukund Rangamani, Maria Rodriquez, Tarun Sharma, David Simmons-Duffin, Piotr Surowka,
 Cumrun Vafa, Amos Yarom and Ho-Ung Yee for extremely useful discussions on ideas presented in this paper. 

The author would like to thank
\textbf{3rd O'Raifeartaigh Conference on Symmetry and Integrability} at the 
Arnold Sommerfeld Center for Theoretical Physics,Munich,
\textbf{Conference on P and CP Odd Effects in Hot and Dense Matter 2012} at the 
RIKEN BNL Research centre workshop,
Brookhaven National Laboratory,
\textbf{ICTS workshop on current topics in string theory} at the  
International Centre for Theoretical Sciences (ICTS),
Indian Institute of Science, Bangalore, 
\textbf{Workshop on Holographic Fluids} conducted by 
Universiteit van Amsterdam (UvA),
Amsterdam and the
\textbf{Workshop on Holographic Duality and Condensed Matter Physics (AdSCMT11)} at the
Kavli Institute for Theoretical Physics (KITP), 
University of California Santa Barbara 
for their hospitality while this work was being completed.

I am  supported by the Harvard Society of Fellows through a junior fellowship.
Finally, I would  like to thank various colleagues at the 
society for interesting discussions.

\appendix
   \begin{center}
      {\bf APPENDICES}
    \end{center}

\section{Notes on Bose-Einstein and Ferm-Dirac moments}\label{app:BoseFermiMoments}
Our objective in this appendix is to evaluate various moments 
associated with Bose-Einstein and Fermi-Dirac distributions. Our
focus would be on those moments which evaluate to polynomials of 
temperature and chemical potential - these moments  are relevant 
for anomaly-induced transport coefficients and to thermal helicity. 
This extends a corresponding dicussion in \cite{Loganayagam:2012pz}
for Fermi-Dirac moments.

We begin with the Gibbs-free energy associated with a single-particle energy level of energy $E_p$
and charge $q$ at a temperature $T=1/\beta$ and chemical potential $\mu$
\begin{equation}\label{eq:gDef}
\begin{split}
g^B &\equiv -\frac{1}{\beta} \ln\brk{1-e^{-\beta E_p}}^{-1}\ ,\qquad
g^F_q \equiv -\frac{1}{\beta} \ln\brk{1+e^{-\beta (E_p-q\mu)} }\\
\end{split}
\end{equation}
where the superscripts $\{B,F\}$ refer to the statistics of the particle
under question. We will only be interested in neutral bosonic states.

The aim of this appendix is to prove the following identities
\begin{equation}\label{eq:gIdentity}
\begin{split}
\int_0^\infty \frac{dE_p}{2\pi} \frac{2^{k-1}}{(k-1)!}\prn{\frac{E_p}{2\pi}}^{k-1}g^B 
&=\frac{2\pi}{8} \brk{ \frac{\tau T}{\tan\prn{\tau T}} }_{\tau^{k+1}}\quad \text{for $k$ odd} \\
\int_0^\infty \frac{dE_p}{2\pi} \frac{1}{(k-1)!}\prn{\frac{E_p}{2\pi}}^{k-1}
\brk{g^F_q+(-1)^{k+1} g^F_{-q}}
&=-2\pi \brk{ \frac{\frac{\tau}{2}T}{\sin\prn{\frac{\tau}{2}T}}e^{\tau
\frac{q\mu}{2\pi}} }_{\tau^{k+1}}
\end{split}
\end{equation}
where the subscript $\tau^{k+1}$ indicates the coefficient of  $\tau^{k+1}$ in the Taylor expansion 
in $\tau$.

Via an integration by parts, these identities can be cast in terms of Bose-Einstein and Fermi-Dirac distributions defined by
\begin{equation}\label{eq:fDef}
\begin{split}
f^B &\equiv \frac{\partial g^B}{\partial E_p} = \frac{1}{e^{\beta E_p}-1}\ ,\qquad
f^F_q \equiv \frac{\partial g^F_q}{\partial E_p}= \frac{1}{e^{\beta (E_p-q\mu)}+1}\\
\end{split}
\end{equation}
so that 
\begin{equation}\label{eq:fIdentity}
\begin{split}
\int_0^\infty \frac{dE_p}{2\pi} \frac{2^{k-1}}{k!}\prn{\frac{E_p}{2\pi}}^k f^B
& =-\frac{1}{8} \brk{ \frac{\tau T}{\tan\prn{\tau T}} }_{\tau^{k+1}}\quad \text{for $k$ odd} \\
\int_0^\infty \frac{dE_p}{2\pi} \frac{1}{k!}\prn{\frac{E_p}{2\pi}}^k
\brk{f^F_q+(-1)^{k+1} f^F_{-q}}
&= \brk{ \frac{\frac{\tau}{2}T}{\sin\prn{\frac{\tau}{2}T}}e^{\tau
\frac{q\mu}{2\pi}} }_{\tau^{k+1}} 
\end{split}
\end{equation}

To prove these identities it is useful to introduce the $n^{th}$ Bernoulli polynomial 
$B_n(x)$ - these are polynomials  defined via the following generating function
\begin{equation}\label{eq:BernoulliGen}
\begin{split}
 \brk{\frac{t\ e^{xt}}{e^t-1}}_{t^n} \equiv \frac{B_n(x)}{n!}\ 
\end{split}
\end{equation}
in terms of which the Taylor coefficients appearing above can be explicitly evaluated
\begin{equation}\label{eq:Bernoulli}
\begin{split}
 \brk{ \frac{\tau T}{\tan\prn{\tau T}} }_{\tau^{k+1}} 
&= \frac{1}{(k+1)!}\prn{\frac{2 i}{\beta}}^{k+1} B_{k+1}\prn{0} \quad \text{for $k$ odd} \\
\brk{ \frac{\frac{\tau}{2}T}{\sin\prn{\frac{\tau}{2}T}}e^{\tau
\frac{q\mu}{2\pi}} }_{\tau^{k+1}} 
&= \frac{1}{(k+1)!}\prn{\frac{ i}{\beta}}^{k+1}
B_{k+1}\prn{\frac{1}{2}+\frac{\beta q\mu}{2\pi i}} \\
\end{split}
\end{equation}

After rewriting in terms of Bernoulli polynomials, the expressions for Bose-Einstein integrals follow directly from the
following integral representations for the Bernoulli numbers 
( See for example formulae [\href{http://dlmf.nist.gov/24.7#E2}{$24.7.2$},\ \href{http://dlmf.nist.gov/24.7#E5}{$24.7.5$}] of
the Digital Library of Mathematical Functions )
\begin{equation}
\begin{split}
B_{2j}(0) &=(-1)^{j+1} 4j\int_0^\infty \frac{t^{2j-1}}{e^{2\pi t}-1} \\
&= \frac{(-1)^k}{\pi} 2j(2j-1)\int_0^\infty t^{2j-2} \ln\brk{1-e^{-2\pi t}} 
\end{split}
\end{equation}
The identities we want to prove follows by taking $t=\beta E_p$ and $2j=k+1$ with $k$ odd.
The Fermi-Dirac identities stated above were proved in \cite{Loganayagam:2012pz} - their proof follows from writing the Fermi-Dirac integrals in terms of
poly-logarithms and then using the following identity (which is a special case of the relation between poly-logarithms and 
Hurwitz zeta function [\href{http://dlmf.nist.gov/25.12#E13}{$25.12.13$},\ \href{http://dlmf.nist.gov/25.11#E14}{$25.11.14$}])
\begin{equation}
\begin{split}
\text{Li}_{k+1}\prn{-e^{\beta q\mu}} +(-1)^{k+1}\text{Li}_{k+1}\prn{-e^{-\beta q\mu}} &=- \frac{(2\pi i)^{k+1}}{(k+1)!}B_{k+1}\prn{\frac{1}{2}+\frac{\beta q\mu}{2\pi i}}
\end{split}
\end{equation}
This completes our proof. In the rest of this appendix we will write down the explicit forms of teh identities stated in
eqns.\eqref{eq:gIdentity} and \eqref{eq:fIdentity}.

The Bosonic integrals are 
\begin{equation}\label{eq:BosonicExp}
\begin{split}
 \int_0^\infty \frac{dE_p}{2\pi}  g^B &= -2\pi\brk{\frac{T^2}{4!}} \\
 \int_0^\infty \frac{dE_p}{2\pi}  \frac{2^2}{2!} \prn{\frac{E_p}{2\pi}}^{2} g^B
&= -2\pi\brk{ 2\frac{T^4}{6!}}\\
 \int_0^\infty \frac{dE_p}{2\pi}  \frac{2^4}{4!}\prn{\frac{E_p}{2\pi}}^{4} g^B
&= -2\pi\brk{ \frac{32}{3}\frac{T^6}{8!}}\\
 \int_0^\infty \frac{dE_p}{2\pi} \prn{\frac{E_p}{2\pi}} f^B &=  \frac{T^2}{4!} \\
 \int_0^\infty \frac{dE_p}{2\pi}  \frac{2^2}{3!} \prn{\frac{E_p}{2\pi}}^{3} f^B
&=  2\frac{T^4}{6!}\\
 \int_0^\infty \frac{dE_p}{2\pi}  \frac{2^4}{5!}\prn{\frac{E_p}{2\pi}}^{5} f^B
&= \frac{32}{3}\frac{T^6}{8!}\\
\end{split}
\end{equation}

Similarly, we can write the identities for the Ferm-Dirac integrals
\begin{equation}\label{eq:gfPolyExp}
\begin{split}
\int_0^\infty \frac{dE_p}{2\pi}   \brk{g^F_q+ g^F_{-q}}
&=-2\pi\brk{\frac{1}{2!}\prn{\frac{q\mu}{2\pi}}^2+\frac{T^2}{4!}} \\
\int_0^\infty \frac{dE_p}{2\pi}  \prn{\frac{E_p}{2\pi}} \brk{g^F_q- g^F_{-q}} &= -2\pi\brk{
{\frac{1}{3!}\prn{\frac{q\mu}{2\pi}}^3+ 
 \prn{\frac{q\mu}{2\pi}}\frac{T^2}{4!}  }  } \\
\int_0^\infty \frac{dE_p}{2\pi} \frac{1}{2!} \prn{\frac{E_p}{2\pi}}^2 \brk{g^F_q + g^F_{-q}} &= -2\pi\brk{
{\frac{1}{4!}\prn{\frac{q\mu}{2\pi}}^4+\frac{1}{2!}\prn{\frac{q\mu}{2\pi}}^2
\frac{ T^2}{4!}+\frac{7}{8}\frac{T^4}{6!}} } \\
\int_0^\infty \frac{dE_p}{2\pi}  \frac{1}{3!} \prn{\frac{E_p}{2\pi}}^3 \brk{g^F_q - g^F_{-q}} &= -2\pi\brk{
{\frac{1}{5!}\prn{\frac{q\mu}{2\pi}}^5+\frac{1}{3!}\prn{\frac{q\mu}{2\pi}}^3
\frac{ T^2}{4!}+\prn{\frac{q\mu}{2\pi}} \frac{7}{8}\frac{T^4}{6!}} } \\
\int_0^\infty \frac{dE_p}{2\pi}  \frac{1}{4!} \prn{\frac{E_p}{2\pi}}^4 \brk{g^F_q + g^F_{-q}} &= -2\pi\brk{
{\frac{1}{6!}\prn{\frac{q\mu}{2\pi}}^6+\frac{1}{4!}\prn{\frac{q\mu}{2\pi}}^4
\frac{ T^2}{4!}+\frac{1}{2!}\prn{\frac{q\mu}{2\pi}}^2
\frac{7}{8}\frac{T^4}{6!}+\frac{31}{24}\frac{T^6}{8!}} } \\
\end{split}
\end{equation}
and
\begin{equation}\label{eq:ffPolyExp}
\begin{split}
\int_0^\infty \frac{dE_p}{2\pi}  \brk{f^F_q- f^F_{-q}} &=\prn{\frac{q\mu}{2\pi}}\\
\int_0^\infty \frac{dE_p}{2\pi}  \prn{\frac{E_p}{2\pi}} \brk{f^F_q+ f^F_{-q}}
&=\frac{1}{2!}\prn{\frac{q\mu}{2\pi}}^2+\frac{T^2}{4!} \\
\int_0^\infty \frac{dE_p}{2\pi}  \frac{1}{2!}\prn{\frac{E_p}{2\pi}}^2 \brk{f^F_q- f^F_{-q}} &=
{\frac{1}{3!}\prn{\frac{q\mu}{2\pi}}^3+ 
 \prn{\frac{q\mu}{2\pi}}\frac{T^2}{4!}  } \\
\int_0^\infty \frac{dE_p}{2\pi} \frac{1}{3!} \prn{\frac{E_p}{2\pi}}^3 \brk{f^F_q + f^F_{-q}} &=
{\frac{1}{4!}\prn{\frac{q\mu}{2\pi}}^4+\frac{1}{2!}\prn{\frac{q\mu}{2\pi}}^2
\frac{ T^2}{4!}+\frac{7}{8}\frac{T^4}{6!}}\\
\int_0^\infty \frac{dE_p}{2\pi} \frac{1}{4!} \prn{\frac{E_p}{2\pi}}^4 \brk{f^F_q - f^F_{-q}} &=
{\frac{1}{5!}\prn{\frac{q\mu}{2\pi}}^5+\frac{1}{3!}\prn{\frac{q\mu}{2\pi}}^3
\frac{ T^2}{4!}+\prn{\frac{q\mu}{2\pi}} \frac{7}{8}\frac{T^4}{6!}}\\
\int_0^\infty \frac{dE_p}{2\pi} \frac{1}{5!} \prn{\frac{E_p}{2\pi}}^5 \brk{f^F_q + f^F_{-q}} &=
{\frac{1}{6!}\prn{\frac{q\mu}{2\pi}}^6+\frac{1}{4!}\prn{\frac{q\mu}{2\pi}}^4
\frac{ T^2}{4!}+\frac{1}{2!}\prn{\frac{q\mu}{2\pi}}^2
\frac{7}{8}\frac{T^4}{6!}+\frac{31}{24}\frac{T^6}{8!}}\\
\end{split}
\end{equation}

\section{Anomaly Polynomials}\label{app:anom}
One of the most elegant ways to encode the anomalies in a system is via the
anomaly polynomial $\mathcal{P}_{anom}(F,\mathfrak{R})$ of gauge field
strengths $F$ and the spacetime curvature $\mathfrak{R}$ . Anomaly polynomials 
are well-covered in various textbooks \cite{Weinberg:1996kr,Bertlmann:1996xk,Bastianelli:2006rx} and lecture 
notes \cite{Harvey:2005it,Bilal:2008qx}. These references however
employ different sign conventions and to avoid confusion we will
spell out our sign conventions in this appendix.

Given a quantum field theory exhibiting anomalies 
in $d=2n$ spacetime dimensions, one can always construct a non-anomalous
theory in one dimension higher by coupling it to an appropriate 
Chern-Simons form $I^{CS}_{2n+1}[A,\Gamma]$ formed out of the connection 1-forms.

If $W_{QFT}$ is the generating function for connected green functions 
for the field theory under question, there exists a  $I^{CS}_{2n+1}$ such that
\begin{equation}\label{eq:CSdef}
  W_{QFT}[\partial\mathcal{M}_{2n+1}] + \int_{\mathcal{M}_{2n+1}} I^{CS}_{2n+1}[A,\Gamma]
\end{equation}
is invariant under gauge and differomorphism transformations of $\partial\mathcal{M}_{2n+1}$ .
We define anomaly polynomial of this field theory to be the $2n+2$ form
\begin{equation}\label{eq:PAnomdef}
\mathcal{P}_{anom}(F,\mathfrak{R}) \equiv dI^{CS}_{2n+1}[A,\Gamma]
\end{equation} 
This then completely fixes the sign conventions for covariant and consistent
currents/anomalies - we will refer the reader to \cite{Jensen:2012kj} for
covariant and consistent currents/anomalies in this sign convention.

We now proceed to present the explicit contibution from chiral matter 
to the anomaly polynomial. To this end, let us begin by defining various
forms relevant for dealing with gravitational anomalies. Let $\mathfrak{R}_{ab}$ be the curvature 2-forms
of the spacetime with
\[ \mathfrak{R}_{ab} \equiv \frac{1}{2!}\ R_{abcd}\ dx^c\wedge dx^d \]
In $d=2n$ dimensions we can think of $\mathfrak{R}_{ab}$ as a real 
$2n\times 2n$ antisymmetric  matrix of 2-forms. At a give point in the
manifold, we can diagonalise it with the diagonal entries being  2-forms
\begin{equation}
 \mathfrak{R}_{ab}=\prn{\begin{array}{cccccc}
+i r_{_1} & 0 &\ldots& \ldots&\ldots& \ldots\\
0 &-i r_{_1} &\ldots & \ldots&\ldots& \ldots \\
\ldots& \ldots&\ldots& \ldots&\ldots& \ldots\\
\ldots& \ldots&\ldots& \ldots&\ldots& \ldots\\
\ldots& \ldots &\ldots & \ldots & +ir_n &0 \\
\ldots& \ldots &\ldots & \ldots & 0 & -ir_n \\
\end{array}}_{ab}
\end{equation}
where ${r}_{j=1,\ldots,n}$ are real 2-forms. Polynomials in these 2-forms can be used to
construct various other useful forms. The rest of this section basically consists
of various such polynomials, relations between them and their generating functions
etc. We start with the most basic form
\[\mathfrak{R}_k \equiv \frac{1}{2} \mathfrak{R}_{a_1}{}^{a_2}\wedge \mathfrak{R}_{a_2}{}^{a_3} \ldots  \wedge\mathfrak{R}_{a_k}{}^{a_1} =  \frac{1+(-1)^k}{2}\sum_j(ir_j)^k\]
so $\mathfrak{R}_k$ is a is a $2k$-form which is a $k^{th}$-degree polynomial in the curvature 2-forms .
As is evident from the expression above, it is non-zero only when $k$ is even.

The next form which is we will introduce is called the $k^{th}$-Pontryagin class denoted by 
$p_{_k}(\mathfrak{R})$ which is $4k$-form using a specific $2k$-th degree polynomial in 
the curvature 2-forms. It is defined via the relation
\begin{equation}
\begin{split}
\text{det}\brk{1+\frac{\tau}{2\pi}\mathfrak{R}} &= \prod_j\brk{1-\prn{\frac{\tau}{2\pi}ir_j}^2}  \equiv \sum_{k} \tau^{2k}p_{_k} (\mathfrak{R}) \\
\end{split}
\end{equation}
This gives
\begin{equation}
\begin{split}
p_{_1}(\mathfrak{R}) &= -\frac{1}{(2\pi)^2} \mathfrak{R}_2 \\
p_{_2}(\mathfrak{R}) &= -\frac{1}{(2\pi)^4} \brk{\frac{1}{2}\mathfrak{R}_4-\frac{1}{2}\mathfrak{R}_2^2} \\
p_{_3}(\mathfrak{R}) &= -\frac{1}{(2\pi)^6} \brk{\frac{1}{3}\mathfrak{R}_6-\frac{1}{2}\mathfrak{R}_2\mathfrak{R}_4+\frac{1}{6}\mathfrak{R}_2^3} \\
\end{split}
\end{equation}
which can be inverted to get 
\begin{equation}
\begin{split}
\frac{1}{(2\pi)^2} \mathfrak{R}_2 &= -p_{_1}(\mathfrak{R}) \\
\frac{1}{(2\pi)^4}\mathfrak{R}_4 &=-2p_{_2}(\mathfrak{R})+p^2_{_1}(\mathfrak{R}) \\
\frac{1}{(2\pi)^6}\mathfrak{R}_6 &=-3p_{_3}(\mathfrak{R}) + 3 p_{_1}(\mathfrak{R})p_{_2}(\mathfrak{R})-p^3_{_1}(\mathfrak{R})  \\
\end{split}
\end{equation}
We will express all the other polynomials in the basis of either $\mathfrak{R}_k$s or $p_{_k}(\mathfrak{R})$.

An object that plays an important role in the anomalies generated by Weyl fermions
is the  Dirac genus (or A-roof genus) $\hat{A}_k(\mathfrak{R})$  which is defined via
\begin{equation}
\begin{split}
\hat{A}\prn{\tau\mathfrak{R}} &\equiv \text{det}^{1/2}\brk{\frac{\frac{1}{2}\frac{\tau}{2\pi}\mathfrak{R}}{\sin\prn{\frac{1}{2} \frac{\tau}{2\pi}\mathfrak{R}}}} = \prod_j\brk{\frac{\frac{1}{2}\frac{\tau}{2\pi}r_j}{\sinh\prn{\frac{1}{2} \frac{\tau}{2\pi}r_j}}}= \sum_k \tau^{2k} \hat{A}_{_k}(\mathfrak{R}) \\
\end{split}
\end{equation}
which gives
\begin{equation}\label{eq:AExp} 
\begin{split}
\hat{A}_{_1}(\mathfrak{R})&=\frac{1}{(2\pi)^2} \brk{\frac{\mathfrak{R}_2}{4!}} = \frac{1}{4!}\brk{- p_{_1}(\mathfrak{R})} \\
\hat{A}_{_2}(\mathfrak{R})&=\frac{1}{(2\pi)^4} \brk{\frac{1}{4}\prn{\frac{\mathfrak{R}_4}{6!}}+\frac{1}{2}\prn{\frac{\mathfrak{R}_2}{4!}}^2}
=\frac{1}{6!(2\pi)^4}\frac{1}{8}\prn{2\mathfrak{R}_4+5\mathfrak{R}_2^2}\\
&=\frac{1}{6!} \brk{-\frac{1}{2}p_{_2}(\mathfrak{R})+\frac{7}{8}p^2_{_1}(\mathfrak{R})}\\
\hat{A}_{_3}(\mathfrak{R})&=\frac{1}{(2\pi)^6}\brk{
\frac{2}{9}\prn{\frac{\mathfrak{R}_6}{8!}}+ \frac{1}{4}\prn{\frac{\mathfrak{R}_2}{4!}}\prn{\frac{\mathfrak{R}_4}{6!}}
+\frac{1}{6}\prn{\frac{\mathfrak{R}_2}{4!}}^3 }  \\
&= \frac{1}{8!(2\pi)^6}\frac{1}{24}\prn{
\frac{16}{3}\mathfrak{R}_6+14\ \mathfrak{R}_4\mathfrak{R}_2+\frac{35}{3}\mathfrak{R}_2^3
}\\
&= \frac{1}{8!}\brk{
-\frac{2}{3}p_{_3}(\mathfrak{R}) +\frac{11}{6}p_{_1}(\mathfrak{R})p_{_2}(\mathfrak{R})
 -\frac{31}{24}p^3_{_1}(\mathfrak{R}) } \\
\end{split}
\end{equation}
The $k^{th}$-Dirac genus  $\hat{A}_{_k}(\mathfrak{R})$ is hence a $4k$-form using a specific $2k$-th degree polynomial in 
the curvature 2-forms.

An object which plays a similar role in the anomalies generated by self-dual
bosons is the  Hirzebruch genus (or L genus) $\hat{L}_k(\mathfrak{R})$  which is
defined via
\begin{equation}
\begin{split}
-\frac{1}{8}\hat{L}(\mathfrak{R})
&\equiv
-\frac{1}{8}\text{det}^{1/2}\brk{
\frac{\frac{t}{2\pi}\mathfrak{R}}
{\tan\frac{t}{2\pi}\mathfrak{R}}}
 = -\frac{1}{8}
 \prod_j\brk{\frac{\frac{t}{2\pi}r_j}{\tanh\prn{\frac{t}{2\pi}r_j}}}
 =-\frac{1}{8}
 \sum_k t^{2k} \hat{L}_k(\mathfrak{R})  \\
\end{split}
\end{equation}
which gives
\begin{equation}
\begin{split}
-\frac{1}{8}\hat{L}_{_1}(\mathfrak{R})&=\frac{1}{(2\pi)^2}
\brk{\frac{\mathfrak{R}_2}{4!}} = \frac{1}{4!}\brk{- p_{_1}(\mathfrak{R})} \\
-\frac{1}{8}\hat{L}_{_2}(\mathfrak{R})&=\frac{1}{(2\pi)^4}
\brk{7\prn{\frac{\mathfrak{R}_4}{6!}}-4\prn{\frac{\mathfrak{R}_2}{4!}}^2}=\frac{
1}{6!(2\pi)^4}\brk{7\mathfrak{R}_4-5\mathfrak{R}_2^2}\\
&=\frac{1}{6!} 2\brk{-7p_{_2}(\mathfrak{R})+p^2_{_1}(\mathfrak{R})}\\
-\frac{1}{8}\hat{L}_{_3}(\mathfrak{R})&=\frac{1}{(2\pi)^6}32\brk{
\frac{31}{9}\prn{\frac{\mathfrak{R}_6}{8!}}-
\frac{7}{4}\prn{\frac{\mathfrak{R}_2}{4!}}\prn{\frac{\mathfrak{R}_4}{6!}}
+\frac{1}{3}\prn{\frac{\mathfrak{R}_2}{4!}}^3 }  \\
&=\frac{1}{8!(2\pi)^6}\frac{8}{3}\brk{
\frac{124}{3}\mathfrak{R}_6-49\mathfrak{R}_4\mathfrak{R}_2+\frac{35}{3}\mathfrak
{R}_2^3
}\\
&= \frac{1}{8!}32\brk{
-\frac{31}{3}p_{_3}(\mathfrak{R})
+\frac{13}{6}p_{_1}(\mathfrak{R})p_{_2}(\mathfrak{R})
 -\frac{1}{3}p^3_{_1}(\mathfrak{R}) } \\
\end{split}
\end{equation}

In case of gravitino, we define
\begin{equation}
\begin{split}
\hat{A}^{(3/2)}(\mathfrak{R})
&\equiv
\brk{\text{tr}\exp\brk{\frac{t}{2\pi}i\mathfrak{R}}-1}\text{det}^{1/2}\brk{
\frac{\frac{1}{2}\frac{t}{2\pi}\mathfrak{R}}{\sin\frac{1}{2}
\frac{t}{2\pi}\mathfrak{R}}}\\
& = \brk{ 2\sum_{j'} \cosh\prn{\frac{t}{2\pi}r_{j'}} -1}
 \prod_j\brk{\frac{\frac{1}{2}\frac{t}{2\pi}r_j}{\sinh\prn{\frac{1}{2}\frac{t}{2\pi}r_j}}}\\
& = \sum_k t^{2k} \hat{A}^{(3/2)}_k(\mathfrak{R})  \\
\end{split}
\end{equation}

The anomaly polynomial for chiral gravitino is given by
\begin{equation}
\begin{split}
\hat{A}^{(3/2)}_{_0}(\mathfrak{R})&= 2n-1\\
\hat{A}^{(3/2)}_{_1}(\mathfrak{R})&=\frac{(2n-25)}{(2\pi)^2} \brk{\frac{\mathfrak{R}_2}{4!}} = \frac{(2n-25)}{4!}\brk{- p_{_1}(\mathfrak{R})} \\
\hat{A}^{(3/2)}_{_2}(\mathfrak{R})&=\frac{1}{(2\pi)^4}\brk{\frac{1}{4}(2n+239)\prn{\frac{\mathfrak{R}
_4}{6!}}+\frac{1}{2}(2n-49)\prn{\frac{\mathfrak{R}_2}{4!}}^2}\\
&=\frac{1}{6!(2\pi)^4}\frac{5}{8}\brk{\frac{2}{5}(2n+239)\mathfrak{R}
_4+(2n-49)\mathfrak{R}_2^2}\\
&=\frac{1}{6!}\brk{-\frac{1}{2}(2n+239)\ p_{_2}(\mathfrak{R})
+\frac{1}{8}(14n+233)\ p^2_{_1}(\mathfrak{R})}\\
\hat{A}^{(3/2)}_{_3}(\mathfrak{R})&=\frac{1}{8!(2\pi)^6}\frac{7}{36}\brk{
\frac{8}{7}(2n-505)\mathfrak{R}_6+3(2n+215)\mathfrak{R}_4\mathfrak{R}_2+\frac{5}{2}(2n-73)\mathfrak{
R}_2^3
}\\
&=\frac{1}{8!}\frac{1}{24}\brk{-16(2n-505)
p_{_3}(\mathfrak{R}) +4(22n-515)p_{_1}(\mathfrak{R})p_{_2}(\mathfrak{R})
-(62n-535)p^3_{_1}(\mathfrak{R})
}\\
\end{split}
\end{equation}
These expressions are to be understood as follows - when these forms say appear in 
the anomaly polynomial of a 4d-theory we need to put $2n=4$ and so on.

The expressions above are valid in free-theories (and interacting theories which are continuously connected to 
free fixed points ). Another class of theories which we will be interested in are CFTs in $d=2n$ spacetime
dimensions which are implicitly defined via AdS/CFT duality through a  gravitational theory in $2n+1$ dimensions.
In this case, we have
\[ W_{CFT}[\partial\mathcal{M}_{2n+1}] = W_{Grav}[\mathcal{M}_{2n+1}] \]
where according to the standard AdS/CFT dictionary, the sources in the $CFT$ path integral
are identified with the non-normalisable modes of the fields in the gravitational side.

We will now sketch the derivation of the anomaly polynomial of the CFT from the dual gravitational theory.
The anomalies of the CFT are reflected in the Chern-Simons term which appears in the gravitational action.
Let us denote this term in the gravitational Lagrangian by $\mathcal{L}^{(AdS)}_{CS}[\hat{A},\hat{\Gamma}]$
where $\{\hat{A},\hat{\Gamma}\}$ are the bulk fields  whose normalisable modes are integrated over while 
performing the gravitational path-integral. This Chern-Simons term leads to boundary anomalies in the gravitational
path-integral,i.e., the gravitational path-integral transforms anomalously under variation of the non-normalisable modes. 

This boundary anomaly can be canceled by adding to the gravitational action
a topological sector of \emph{non-dynamical} bulk metric and gauge-fields 
with an action given by 
\[ W_{Topo}= -\int_{\mathcal{M}_{2n+1}}\mathcal{L}^{(AdS)}_{CS}[A,\Gamma] \]
where the non-normalisable part of the non-dynamical fields $\{A,\Gamma\}$ are taken to be identical to the
non-normalisable part of the dynamical fields $\{\hat{A},\hat{\Gamma}\}$. Thus
we conclude that the functional
\[  W_{Grav}[\mathcal{M}_{2n+1}]-\int_{\mathcal{M}_{2n+1}}\mathcal{L}^{(AdS)}_{CS}[A,\Gamma]=W_{CFT}[\partial\mathcal{M}_{2n+1}]-\int_{\mathcal{M}_{2n+1}}\mathcal{L}^{(AdS)}_{CS}[A,\Gamma] \]
is invariant under boundary transformations. Comparing this functional against equations \eqref{eq:CSdef}
and \eqref{eq:PAnomdef} , we conclude that the anomaly polynomial of the dual CFT is given by 
\begin{equation}
\label{eq:PAnomAdSCFT} \mathcal{P}^{CFT}_{anom}(F,\mathfrak{R}) = -d\mathcal{L}^{(AdS)}_{CS}[A,\Gamma] 
\end{equation}

The anomaly polynomial for a bunch of free chiral p-forms/fermions/gravitini in a $d=(2n-1)+1$ 
dimensional spacetime is given by 
\begin{equation}
\begin{split}
\prn{\mathcal{P}_{anom}}_{d=2n}&=- 2\pi\left[-\frac{1}{8}\sum_{\text{Chiral forms}}  \chi_{_{d=2n}}\ \hat{L}\prn{\tau \mathfrak{R}}\ 
+\sum_{\text{Weyl species}}  \chi_{_{d=2n}}\ \hat{A}\prn{\tau \mathfrak{R}}\ 
e^{\frac{\tau}{2\pi}qF}\right.\\
&\qquad \left.+\sum_{\text{Weyl gravitino species}}  \chi_{_{d=2n}}\ \hat{A}^{(3/2)}\prn{\tau \mathfrak{R}}\ 
e^{\frac{\tau}{2\pi}qF} \right]_{\tau^{n+1}}
\end{split}
\end{equation}
where we have assumed  that there is a single $U(1)$ symmetry under which the charge of the
of a particle is denoted by the letter $q$. 

The symbol $\chi_{_{d=2n}}$ denotes chirality in 
$d=2n$ dimensions : our convention for chirality is as follows - given a particular irreducible 
chiral representation of Poincare group (definining a single particle) we can always choose 
a unique state in that representation with  the spin along $x^{2k-1},x^{2k}$ plane $\sigma_k$ 
being maximum/positive for $k=1,\ldots, n-1$ (this is
the highest weight state of the little group $so(2n-2)$). We will define the 
chirality of a  representation $\chi_{_{d=2n}} =+1$ if the momentum along $x^{2n-1}$ axis of this highest weight 
state is positive. And $\chi_{_{d=2n}} =-1$ if the momentum along $x^{2n-1}$ axis of this highest weight 
state is negative. Further the sum in the formula above is over the species which means that 
each particle/anti-particle pair contributes one term to the sum (the answer does not depend on
whether we take the chirality/charge of a particle or the anti-particle ).

In conformal field theories with sufficient supersymmetry, there is often a direct relation between the central
charges appearing in the Weyl anomaly and the anomaly polynomial for the R-charge. For example, we have
\begin{equation}
\begin{split}
\prn{\mathcal{P}_{anom}}^{CFT}_{d=2} &= -2\pi\brk{
k_R\prn{\frac{F_R}{2\pi}}^2 -k_L\prn{\frac{F_L}{2\pi}}^2
- p_{_1}(\mathfrak{R}) \frac{c_R-c_L}{24}} \\
\prn{\mathcal{P}_{anom}}^{\mathcal{N}=(0,2)\ \text{SCFT}}_{d=2} &= -2\pi\brk{ \frac{c_R}{3}\frac{1}{2!}\prn{\frac{F_R}{2\pi}}^2
 -\frac{c_R-c_L}{4!}p_{_1}(\mathfrak{R}) }\\
\prn{\mathcal{P}_{anom}}^{\mathcal{N} = 1\ \text{SCFT}}_{d=4} &= -2\pi \brk{\frac{2^4}{3}\prn{\frac{5}{3}a- c} \frac{1}{3!}\prn{\frac{F_R}{2\pi}}^3
 -\frac{2^4(a-c)}{4!}p_{_1}(\mathfrak{R})\prn{\frac{F_R}{2\pi}} }
\end{split}
\end{equation}

We now proceed to write down the explicit expressions for $\mathcal{P}_{anom}$ by using the formula above.
\begin{equation}\label{eq:PExp2d4d6d}
\begin{split}
(\mathcal{P}_{anom})_{_{d=1+1}} &= -2\pi \sum_{species} \chi_{_{d=2}} \brk{\frac{1}{2!}\prn{\frac{F}{2\pi}}^2\prn{q^2_{1/2}+q^2_{3/2}}
-\frac{1}{4!}p_{_1}(\mathfrak{R}) \prn{1_0+1_{1/2} -23_{3/2}} }\\
(\mathcal{P}_{anom})_{_{d=3+1}} &= -2\pi \sum_{species} \chi_{_{d=4}}\brk{\frac{1}{3!}\prn{\frac{F}{2\pi}}^3\prn{q^3_{1/2}+3\ q^3_{3/2}}- 
 \frac{1}{4!}p_{_1}(\mathfrak{R})\prn{\frac{F}{2\pi}}\prn{q_{1/2} -21 q_{3/2}} }  \\
 (\mathcal{P}_{anom})_{_{d=5+1}}&= -2\pi\sum_{species}  \chi_{_{d=6}}  \left[\frac{1}{4!}\prn{\frac{F}{2\pi}}^4\prn{q^4_{1/2}+5\ q^4_{3/2}}
-  \frac{1}{4!}p_{_1}(\mathfrak{R})\frac{1}{2!}\prn{\frac{F}{2\pi}}^2 \prn{q^2_{1/2} -19 q^2_{3/2}}\right.\\
&\qquad\left.-\frac{1}{6!}p_{_2}(\mathfrak{R})\frac{1}{2}\prn{28_{2}+1_{1/2}+245_{3/2}}
+\frac{1}{6!}p^2_{_1}(\mathfrak{R})\frac{1}{8}\prn{16_{2}+7_{1/2}+275_{3/2}}\right]\\
(\mathcal{P}_{anom})_{_{d=7+1}} &= -2\pi\sum_{species}  \chi_{_{d=8}}  \left[\frac{1}{5!}\prn{\frac{F}{2\pi}}^5\prn{q^5_{1/2}+7\ q^5_{3/2}}
-  \frac{1}{4!}p_{_1}(\mathfrak{R})\frac{1}{3!}\prn{\frac{F}{2\pi}}^3 \prn{q^3_{1/2} -17 q^3_{3/2}}\right.\\
&\qquad\left.-\frac{1}{6!}p_{_2}(\mathfrak{R})\prn{\frac{F}{2\pi}}\frac{1}{2}\prn{q_{1/2}+247 q_{3/2}}
+\frac{1}{6!}p^2_{_1}(\mathfrak{R})\prn{\frac{F}{2\pi}}\frac{1}{8}\prn{7 q_{1/2}+289 q_{3/2}}\right]\\
(\mathcal{P}_{anom})_{_{d=9+1}} &= -2\pi\sum_{species}  \chi_{_{d=8}}  \left[\frac{1}{6!}\prn{\frac{F}{2\pi}}^6\prn{q^6_{1/2}+9\ q^6_{3/2}}
-  \frac{1}{4!}p_{_1}(\mathfrak{R})\frac{1}{4!}\prn{\frac{F}{2\pi}}^4 \prn{q^4_{1/2} -15 q^3_{3/2}}\right.\\
&\qquad\left.-\frac{1}{6!}p_{_2}(\mathfrak{R})\frac{1}{2!}\prn{\frac{F}{2\pi}}^2\frac{1}{2}\prn{q^2_{1/2}+249\ q^2_{3/2}}
+\frac{1}{6!}p^2_{_1}(\mathfrak{R})\frac{1}{2!}\prn{\frac{F}{2\pi}}^2\frac{1}{8}\prn{7 q^2_{1/2}+303\ q^2_{3/2}}\right.\\
&\qquad\left.-\frac{1}{8!}p_{_3}(\mathfrak{R})\frac{2}{3}\prn{496_{4}+1_{1/2}-495_{3/2}}
+\frac{1}{8!}p_{_2}(\mathfrak{R})p_{_1}(\mathfrak{R})\frac{1}{6}\prn{416_4+ 11_{1/2}-405_{3/2}}\right.\\
&\qquad\left.\qquad-\frac{1}{8!}p^3_{_1}(\mathfrak{R})\frac{1}{24}\prn{256_{4}+31_{1/2}-225_{3/2}}
\right]\\
\end{split}
\end{equation}
where we assume there is one U(1) symmetry under which in general both the Weyl fermions and  Weyl gravitini are charged.
The subscript $\{1/2,3/2\}$ denote contributions from a Weyl fermion and a Weyl gravitino respectively (if the fermions
are Weyl Majorana, the contributions have to be halved). The integer subscripts $(\ldots)_p$ denote 
contributions from chiral $p$-forms (with $p+1$-form field strengths)
which are assumed to be neutral under the global symmetry $U(1)$.

We can compare these anomaly polynomials to the transport coefficient $\mathfrak{F}^\omega_{anom}$ derived in 
the main text
\begin{equation}\label{eq:FExp2nd}
\begin{split}
(\mathfrak{F}^\omega_{anom})_{_{d=1+1}} &= -2\pi \sum_{species} \chi_{_{d=2}} \brk{\frac{1}{2!}\prn{\frac{\mu}{2\pi}}^2\prn{q^2_{1/2}+q^2_{3/2}}
+\frac{T^2}{4!}\prn{1_0+1_{1/2} +1_{3/2}} }\\
(\mathfrak{F}^\omega_{anom})_{_{d=3+1}} &= -2\pi \sum_{species} \chi_{_{d=4}}\brk{\frac{1}{3!}\prn{\frac{\mu}{2\pi}}^3\prn{q^3_{1/2}+3\ q^3_{3/2}}+ 
 \frac{T^2}{4!}\prn{\frac{\mu}{2\pi}}\prn{q_{1/2} +3\ q_{3/2}} }  \\
(\mathfrak{F}^\omega_{anom})_{_{d=5+1}}&= -2\pi\sum_{species}  \chi_{_{d=6}}  \left[\frac{1}{4!}\prn{\frac{\mu}{2\pi}}^4\prn{q^4_{1/2}+5\ q^4_{3/2}}
\right.\\
&\qquad\left.+ \frac{T^2}{4!}\frac{1}{2!}\prn{\frac{\mu}{2\pi}}^2 \prn{q^2_{1/2} +5\ q^2_{3/2}}
+\frac{T^4}{6!}\frac{1}{8}\prn{16_{2}+7_{1/2}+35_{3/2}}\right]\\
(\mathfrak{F}^\omega_{anom})_{_{d=7+1}} &= -2\pi\sum_{species}  \chi_{_{d=8}}  \left[\frac{1}{5!}\prn{\frac{\mu}{2\pi}}^5\prn{q^5_{1/2}+7\ q^5_{3/2}}
\right.\\
&\qquad\left.+  \frac{T^2}{4!}\frac{1}{3!}\prn{\frac{\mu}{2\pi}}^3 \prn{q^3_{1/2} +7 q^3_{3/2}}
+\frac{T^4}{6!}\prn{\frac{\mu}{2\pi}}\frac{1}{8}\prn{7 q_{1/2}+49 q_{3/2}}\right]\\
(\mathfrak{F}^\omega_{anom})_{_{d=9+1}} &= -2\pi\sum_{species}  \chi_{_{d=8}}  \left[\frac{1}{6!}\prn{\frac{\mu}{2\pi}}^6\prn{q^6_{1/2}+9\ q^6_{3/2}}
+ \frac{T^2}{4!}\frac{1}{4!}\prn{\frac{\mu}{2\pi}}^4 \prn{q^4_{1/2} +9 q^3_{3/2}}\right.\\
&\qquad\left.+\frac{T^4}{6!}\frac{1}{2!}\prn{\frac{\mu}{2\pi}}^2\frac{1}{8}\prn{7 q^2_{1/2}+63\ q^2_{3/2}}
+\frac{T^6}{8!}\frac{1}{24}\prn{256_{4}+31_{1/2}+279_{3/2}}
\right]\\
\end{split}
\end{equation}

\begin{equation}\label{eq:anomFPCorr}
\begin{split}
\mathfrak{F}_{anom}^\omega[T,\mu] = \mathcal{P}_{anom} \brk{ F \mapsto \mu, p_1(\mathfrak{R}) \mapsto - T^2 , p_{k>1}(\mathfrak{R}) \mapsto 0 } 
+\Delta \mathfrak{F}^\omega_{anom}
\end{split}
\end{equation}
where
\begin{equation}\label{eq:deltaF}
\begin{split}
(\Delta \mathfrak{F}^\omega_{anom})_{_{d=1+1}} &= -2\pi T^2 \sum_{species} \chi_{_{d=2}} 1_{3/2} \\
(\Delta \mathfrak{F}^\omega_{anom})_{_{d=3+1}} &= -2\pi T^2 \sum_{species} \chi_{_{d=4}}
 \prn{\frac{\mu}{2\pi}} q_{3/2}   \\
(\Delta \mathfrak{F}^\omega_{anom})_{_{d=5+1}}&= -2\pi T^2\sum_{species}  \chi_{_{d=6}}  \left[\frac{1}{2!}\prn{\frac{\mu}{2\pi}}^2  q^2_{3/2}
-\frac{T^2}{4!} 1_{3/2}\right]\\
(\Delta \mathfrak{F}^\omega_{anom})_{_{d=7+1}} &= -2\pi T^2 \sum_{species}  \chi_{_{d=8}}  \left[
\frac{1}{3!}\prn{\frac{\mu}{2\pi}}^3  q^3_{3/2}
-\frac{T^2}{4!}\prn{\frac{\mu}{2\pi}} q_{3/2}\right]\\
(\Delta \mathfrak{F}^\omega_{anom})_{_{d=9+1}} &= -2\pi T^2 \sum_{species}  \chi_{_{d=8}}  \left[\frac{1}{4!}\prn{\frac{\mu}{2\pi}}^4 
 q^3_{3/2}
-\frac{T^2}{4!}\frac{1}{2!}\prn{\frac{\mu}{2\pi}}^2 q^2_{3/2}+\frac{T^4}{8!}21_{3/2} \right]\\
\end{split}
\end{equation}


\bibliographystyle{JHEP}
\bibliography{chiralZ}

\end{document}